\documentclass{aa}

\usepackage{newtxtext,newtxmath}
\usepackage[T1]{fontenc}
\usepackage{ae,aecompl}
\usepackage[english]{babel}
\usepackage[usenames]{xcolor}
\usepackage{mathtools}
\usepackage{hyperref}
\usepackage{url}
\usepackage{soul}

\usepackage{graphicx}   % Including figure files
\usepackage{amsmath}    % Advanced maths commands
\usepackage{amssymb}    % Extra maths symbols
\usepackage{adjustbox}
\usepackage{threeparttable} %Footnote in tables
\usepackage[normalem]{ulem}
\begin{document}

\title{Dust trapping around Lagrangian points in protoplanetary disks}

   \author{Mat\'ias Montesinos
          \inst{1,2,3}\fnmsep\thanks{E-mail: matias.montesinos@uv.cl},
          Juan Garrido-Deutelmoser\inst{4,3},
          Johan Olofsson\inst{1,3},
          Cristian A. Giuppone\inst{5},
          Jorge Cuadra\inst{6,3},
          Amelia Bayo\inst{1,3}, 
          Mario Sucerquia\inst{1,3},
          Nicol\'as Cuello\inst{7}
          }

   \institute{Instituto de F\'isica y Astronom\'ia, Universidad de Valpara\'iso, Chile
              \and
             Chinese Academy of Sciences South America Center for Astronomy, National Astronomical Observatories, CAS, Beijing 100012, China
        \and
        N\'ucleo Milenio de Formaci\'on Planetaria (NPF), Chile
         \and
        Instituto de Astrof\'isica, Pontificia Universidad Cat\'olica de Chile, Santiago, Chile
        \and
        Universidad Nacional de C\'ordoba, Observatorio Astron\'omico - IATE. Laprida 854, 5000 C\'ordoba, Argentina
        \and
        Departamento de Ciencias, Facultad de Artes Liberales, Universidad Adolfo Ib\'a\~nez, Av.\ Padre Hurtado 750, Vi\~na del Mar, Chile
        \and
        Univ. Grenoble Alpes, CNRS, IPAG, F-38000 Grenoble, France
             }

% These dates will be filled out by the publisher
\date{}

% Abstract of the paper
\abstract
  % context heading (optional)
  {}
  % {} leave it empty if necessary
  % aims heading (mandatory)
    {Trojans are defined as objects that share the orbit of a planet at the stable Lagrangian points $L_4$ and $L_5$. In the Solar System, these bodies show a broad size distribution ranging from micrometer($\mu$m) to centimeter(cm) particles (Trojan dust) and up to kilometer (km) rocks (Trojan asteroids). It has also been theorized that earth-like Trojans may be formed in extra-solar systems. The Trojan formation mechanism is still under debate, especially theories involving the effects of dissipative forces from a viscous gaseous environment.}
    % methods heading (mandatory)
    {We perform hydro-simulations to follow the evolution of a protoplanetary disk with an embedded  1--10 Jupiter-mass planet. On top of the gaseous disk, we set a distribution of $\mu$m--cm dust particles interacting with the gas. This allows us to follow dust dynamics as solids get trapped around the Lagrangian points of the planet.}
% results heading (mandatory)
    {We show that large vortices generated at the Lagrangian points are responsible for dust accumulation, where the leading Lagrangian point $L_4$ traps a larger amount of submillimeter (submm) particles than the trailing $L_5$, which traps mostly mm--cm particles. However, the total bulk mass, with typical values of $\sim M_{\rm moon}$, is more significant in $L_5$ than in $L_4$, in contrast to what is observed in the current Solar System a few gigayears later. Furthermore, the migration of the planet does not seem to affect the reported asymmetry between $L_4$ and $L_5$.}
% conclusions heading (optional), leave it empty if necessary 
    {The main initial mass reservoir for Trojan dust lies in the same co-orbital path of the planet, while dust migrating from the outer region (due to drag) contributes very little to its final mass, imposing strong mass constraints for the in situ formation scenario of Trojan planets.}

% Select between one and six entries from the list of approved keywords.
% Don't make up new ones.

\titlerunning{Dust trapping around Lagrangian points}
\authorrunning{Mat\'ias Montesinos et al.}
\maketitle
%%%%%%%%%%%%%%%%%%%%%%%%%%%%%%%%%%%%%%%%%%%%%%%%%%

%%%%%%%%%%%%%%%%% BODY OF PAPER %%%%%%%%%%%%%%%%%%

\section{Introduction}

In 1772 Joseph-Louis Lagrange identified five equilibrium points ($L_1$, $L_2$, $L_3$, $L_4$, and $L_5$) derived from the restricted three-body problem \citep{Lagrange1772}, in which a particle of negligible mass orbits under the action of two massive bodies (e.g., a star--planet system). Two of these points, $L_4$ and $L_5$, lie in the orbit of the smaller body (planet), each one at the vertex of an equilateral triangle with the opposing joint base formed by the line of the two massive bodies. $L_4$ is located $\simeq +\pi/3$ rad at the leading position of the planet, while $L_5$ is located at the trailing co-orbital region at $\simeq -\pi/3$ rad (with respect to the planet).

The geometry of $L_4$ and $L_5$ implies that the ratio of their distances to the barycenter is equal to the ratio of the two masses. Consequently, the net gravitational force from the planet--star system is zero at these locations. Hence, $L_4$ and $L_5$ should be linearly stable under small perturbations. \cite{Gascheau1843} determined that for sufficiently small ratios ($<1/27$) of the star--planet mass system, it should accumulate nonmassive objects called Trojans (see \citealt{Brouwer+1961, Szebehely1967, Sosnitskii1996} for more details on the stability). The first Jovian Trojan (588 Achilles) was discovered by Max Wolf in 1906 \citep{Wolf1907}. Since then, thousands have been reported\footnote{\url{https://www.minorplanetcenter.net/iau/lists/JupiterTrojans.html}}.  More recently, the first Neptune Trojan asteroids 2001 QR322 and 2008 LC15 \citep{Sheppard+2010} were discovered.

Several efforts have been made to understand the dynamics of Trojans and their origins. For instance, \cite{Morbidelli+2005} suggest that Trojan asteroids could have been formed in distant regions to be later captured into co-orbital motion during the migration of the giant planet in the context of the Nice model, where the evolution of the planets is followed after the gas disk has dissipated \citep{Tsiganis+2005}.

The origin of Trojans could also be  connected to an even earlier stage of the Solar System, where there is a gas-rich environment. 
Several hydrodynamical simulations  
indeed show that asymmetric overdensities in the gas are produced in co-rotation with the planet, favoring $L_5$ over $L_4$ (e.g., \citealt{Val-Borro+2006, Lyra+2009}). Therefore, the accumulation of Trojan dust should be larger in $L_5$ compared to $L_4$. However, this is contrary to observations of Jovian Trojans, which show 
a ratio of the number of asteroids $N(L_4)/N(L_5) \approx  1.8$ for Trojans with diameters $D > 2\,$ km \citep{Yoshida+2005}.

One possible explanation for this discrepancy is that we now observe the end result of multiple physical processes (e.g., drag forces, collisions, grain growth) that have modified the Trojan population  since formation of the Solar System   
(e.g., \citealt{Milani+2017}). For instance, \citet{DiSisto+Ramos.2019} analyzed the observed Trojan population through numerical simulations and concluded that the Trojan escape rate from $L_5$ in the lifetime of the Solar System is $\sim$ 1.1 times greater than that from $L_4$, and is mainly produced by gravitational interactions with the other planets (from Venus to Neptune).
\cite{Pirani+2019} studied the consequences of planetary migration on the minor bodies of the Solar System through N-body simulations. 
They find that inward migration produces a more populated leading swarm ($L_4$) than the trailing one ($L_5$), in agreement with observations, while a nonmigrating planet results in symmetric swarms. However, the study of these latter authors is limited by a simplified treatment of the drag force that mimics the effect of the gas-phase of the protoplanetary disk. A complementary explanation could be the lack of systematic observations, or bias in them.  

The bias problem concerns two critical aspects. The first is related to the limited amount of time for the observation, which translates to a limitation in the absolute magnitude $H$ that can be reached. For instance, \cite{Lagerkvist+2002} show that observing Trojans to a limit of $H = 11$ mag indicates that the $L_5$ swarm is 75 \% of $L_4$, while down to $H = 13$ mag shows that $L_5$ swarm is 76\% of $L_4$. The other aspect concerns the covered area to observe a Lagrangian swarm. Unfitted estimations of the density area lead to inaccurate population estimations. For instance, \cite{Jewitt+2000} estimate the $L_4$ Trojan population by analyzing an area between $L_4$ and $L_3$, where the distribution is more spread than between $L_4$ and Jupiter. Follow-up observations indicate that their results were overestimated by 40\% \citep{Lagerkvist+2002}. Different inclination distributions of Trojans and low albedo (e.g., \citealt{Yoshida+2005}) also contribute to bias these detections, notably in the case of small Jovian objects with $D \lesssim 1 m$ (e.g., \citealt{Jedicke+2002}). Large correction factors are therefore required to overcome these observational biases \citep{Karlsson+2010}. 

An interesting scenario is the possibility of a co-orbital planet located at a Lagrangian point. In that context, \cite{Lyra+2009} studied the effect of high-pressure regions around the Lagrangian points $L_4$ and $L_5$ of a giant planet. They show that large bodies (m--km) accumulate in tadpole orbits, suggesting the formation of a Trojan Earth-mass planet in situ. This model requires a considerably massive and dense (self-gravitating) disk. In a closely related context, \cite{Cresswell+2009} investigated the evolution of Trojan planets embedded in a gaseous disk, from which they can grow to become gas giant planets. The 2D numerical simulations of these latter authors show than once a Trojan planet is placed to grow, the system is stable for about $\sim 10^9 \rm yrs$. 

Despite the strong assumptions needed to produce high-mass Trojan planets, active searches for these objects in extra-solar
systems are currently taking place. For instance, \cite{Giuppone+2012} analyzed the possible detection of exoplanets in co-orbital motion, namely the TROY project\footnote{\url{https://www.troy-project.com/}}, an observational and theoretical effort to understand the evolution of planetary systems from the characterization of (still not detected)  exo-Trojan planets \citep{Lillo-Box+2018a, Lillo-Box+2018b}. This characterization raises several open questions, such as: are large Trojan asteroids formed in situ (by aggregates of micron-cm particles to km rocks) or captured,  for example during migration? What could be the maximum mass reservoir of dust that can accumulate in Lagrangian points? What is the dynamical evolution of the early dust grains initially present when a protoplanetary object appears? 

In this work, we study the early evolution of Trojan dust with the aim being to understand the current configuration of Trojans  in the Solar System and the possibility of finding Trojan exoplanets. We model the dynamical evolution of the primordial dust present at the early stage of a gaseous protoplanetary disk, with a particular focus on the dust concentration around the Lagrangian points $L_4$ and $L_5$, and its stability on short timescales ($10^4$ yrs). 

The paper is structured as follows: In Section \ref{numerical_model} we provide a description of the numerical setup.  
In Section \ref{physical}, we discuss the physical considerations regarding vortices and Lagrangian points. In Section \ref{results}, we present the results of the gas and dust simulations. A discussion is provided in Section \ref{discussion}, and our main conclusions are drawn in Section \ref{conclusions}.

\section{Numerical model}\label{numerical_model}

We follow the evolution of a dusty, gaseous, viscous self-gravitating protoplanetary disk with an embedded planet. For that purpose, we divide our simulations into two stages; the first computes the gas hydrodynamics alone by solving the Navier-Stokes equation and a nonstationary energy equation using a revised version of the 2D FARGO-ADSG code \citep{Masset-2000, Baruteau-Masset-2008}. In our version, we model a passively heated disk irradiated by the central star. The disk includes a radiative cooling mechanism assuming black-body radiation, with a nonstationary energy equation. For more details, see \cite{montesinos+2016}.
The second stage follows the dust dynamics computed on top of the gaseous stage, where the outputs of the hydro-simulation (first stage) are used as inputs for the dust code (second stage), which computes the dynamics of the Lagrangian particles. A complete description of the dust code and our methodology can be found in \cite{Cuello2019}.

The evolution of the disk is followed for about 10,000 years (or 460 planetary orbits). To check that our results correspond to a stationary regime, we also run a supplementary model running up to 1000 orbits for our fiducial model, obtaining almost the same results. We do not present such a model in this work. A description of the dust and gas stage is outlined in the following section.

\subsection{First phase: gas dynamic setup}\label{Gas_initialization}

Our fiducial model includes a Jupiter-mass planet located at 7.8 au. 
The choice of the planet location was motivated by its generated gap, which may be scalable to some observations such as the cavity reported in HD 100546, which could be carved by a Jupiter-mass planet (e.g., \citealt{Bouwman+2003, Tatulli+2011}).

The orbital period of the planet is about $T_{\rm p} = 21.8$ yr, and its gravitational effect is introduced smoothly by a taper function such that its final mass is reached after $N_{\rm{taper}}=10$ orbits;

\begin{equation}
 M_\textrm{p}(t) = \begin{cases}
               M_\textrm{p}\cdot \sin^2(\frac{\pi}{2}\frac{t}{N_{\rm{taper}}\, T_\textrm{p}})  &       t < 10 \, T_\textrm{p} \\       
               M_\textrm{p}            &        {\rm otherwise.}
      \end{cases}
\end{equation}

The initial density profile of the disk is assumed to be
\begin{equation}
    \Sigma(r) = 8.9 \left( \frac{{\rm au}}{r}  \right) \rm{gr~cm^{-2}},
    \label{eqdens}
\end{equation}
distributed in a radially logarithmic grid of $512~ (\rm radial) \times 1024~  \rm (azimuth)$ sectors with cylindrical coordinates $r$, $\phi$. The inner radius $r_{\rm in}$ of the disk is located at 2.5 au from the central star,  and the outer radius $r_{\rm out}$ at 15 au. From the density profile \eqref{eqdens} and the grid limits, the initial disk mass gives $M_{\rm disk}^{\rm gas} = \int_{r_{\rm in}}^{r_{\rm out}} 2 \pi \Sigma(r) r dr  = 2.4 \times 10^{-3} M_\odot$, which is compatible with typical circumstellar disk masses (e.g., \citealt{Andrews+2005}).

Despite the fact that this mass corresponds to a low disk mass, we include self-gravitating effects in our calculations. These effects could impact the dust dynamics, even for a large Toomre parameter such as that obtained from our disk parameters \citep{Baruteau+2016}.

We solve a nonstationary energy equation, in which the source term is given by stellar irradiation
\begin{equation}
    Q^+_\star = (1 - \beta) \frac{L_\star}{4 \pi r^2} \frac{H}{r} \left( \frac{{\rm d}\ln H}{{\rm d}\ln r} - 1  \right),
\end{equation}
where $\beta$ is the albedo (set to zero), $L_\star$ the stellar luminosity, and $H$ the scale-height computed assuming local hydrostatic equilibrium $H = \frac{c_s}{v_\phi} r$, with $c_s$ and $v_\phi$ being the sound speed and azimuthal velocity, respectively.

The radiative cooling mechanism is given by black-body emission,
\begin{equation}
    Q^- =  \frac{ 2 \sigma_{\rm SB} T^4}{\tau},
\end{equation}
 where $ \sigma_{\rm SB}$ is the Stefan-Boltzmann constant, $\tau = \frac{1}{2} \kappa \Sigma$ the optical depth, and $T$ the midplane temperature. We simplify our calculations by using a constant disk opacity $\kappa = 1~  \rm cm^2 g^{-1}$.  This choice is justified by noting that the 
 Rosseland mean opacity in the temperature and density range of this work gives a mean value $\kappa \sim 1 ~ \rm cm^2 g^{-1}$ \citep{Semenov+2003}.

The disk temperature is initialized by
\begin{equation}
    T(t=0, r) = h^2 \frac{G M_\star}{r} \frac{\mu}{\mathcal{R}},
\end{equation}
where $h = h_0 r^f$ is the aspect ratio of the disk. The factor $f$ is the flare of the disk, which is assumed to be $f = 1/7$, obtained from equating  $Q^+_\star = Q^-$, corresponding to a quasi-steady regime. This choice helps to reach a quasi-steady situation as fast as possible. The initial aspect ratio is set to $h_0 = 0.05$. A complete description of the energy equation and its initialization can be found in \cite{montesinos+2016} and \cite{Cuello2019}. 

To explore other disk configurations we compute different models taking the same density distribution (Eq. \ref{eqdens}), but changing the mass of the planet $M_{\rm p} = \{1, 5\} M_{\rm J}$, the stellar luminosity $L_\star = \{1, 5\} \rm {L_\odot}$, and the disk turbulent viscosity $\alpha = \{10^{-4}, 10^{-3}, 10^{-2}\}$\footnote{available data on protoplanetary disks suggest $\alpha \sim 10^{-4}$, e.g., \cite{Zhang+2018, Flaherty+2020}.}, where $\alpha$ is the turbulent viscosity prescription from \cite{Shakura+Sunyaev1973}. We also run an extra migrating model, where the planet was initially located at 5.2 au.
We summarize the explored parameter space in Table \ref{models_parameters}.

\begin{table}
\begin{center}
\begin{threeparttable}
\begin{tabular}{ccccc}
    \hline
    \hline
    \multicolumn{5}{c}{Model parameters}\\
    \hline
    \hline
    \textbf{Model} & $M_{p\rm }$ [$M_J$] & $r_{\rm p}$ [au] & $L_{\star}$ [$L_\odot$] & $\alpha$  \\ 
     1 & 1 & 7.8 & 5 & $10^{-4}$ \\
     2 & 1 & 7.8 & 5 & $10^{-2}$ \\
     3 & 5 & 7.8 & 5 & $10^{-4}$ \\
     4 & 5 & 7.8 & 1 & $10^{-4}$ \\
     5 & 5 & 7.8 & 1 & $10^{-2}$ \\
     6 & 1 & 7.8 & 5 & $10^{-3}$\\
    \hline

     7\tnote{a} & 10 & 5.2 & 1 & $10^{-1}$\\
     8\tnote{b} & 10 & 5.2 & 1 & $10^{-1}$\\
    \hline
\end{tabular}
\begin{tablenotes}\footnotesize
\item[a] Nonmigrating model 
\item[b] Migrating model 
\end{tablenotes}
\end{threeparttable}
\end{center}
\caption{Model parameters of the hydro-simulations.}
\label{models_parameters}
\end{table}

\subsection{Second phase: dust dynamics setup}

The dust phase is a post-processing calculation where an ensemble of $N$-independent dust particles react to drag forces produced by the gas and gravitational forces from the star, the gaseous disk, and the embedded planet. The particles are described as Lagrangian test particles that do not interact with one another.  The dust code is based on previous work by
\cite{Paardekooper-2007, Zsom+2011}.

The drag force on each particle is computed as

\begin{equation}
    \textbf{F}_{\rm drag} = - \frac{\Omega(r)}{St} \Delta \textbf{v},
\label{eqFdrag}
\end{equation}
where $\Omega(r)$ is the angular velocity of the gas, $\Delta \textbf{v}$ the relative velocity between the gas and the dust, and $St$ the Stokes number, which is computed as an interpolation between Epstein and Stokes regimes, which is valid for small and large particles, respectively (see Appendix C from \citealt{Cuello2019}, and \cite{phdthesis}). 

Although we use a more convenient way to compute the Stokes number, most of the particles in the simulation follow the Epstein regime, which is valid for particles with sizes smaller than 9/4 of their mean free path \citep{Weidenschilling+1977}. In the Epstein regime, the Stokes number for each particle $i$ is given by
\begin{equation}
    {\rm St}_i = \frac{\rho_{\rm d} s_i }{\rho_{\rm g} c_{\rm 
    s}} \Omega(r),
\end{equation}
where $\rho_{\rm d}$ is the bulk density of the dust, $s_i$ the size of the particle $i$, $\rho_{\rm g}$ the local gas density, and $c_{\rm s}$ the local sound speed. The Stokes number for each particle varies depending on the position $r$ of that particle.

In our calculations, we also include turbulent diffusion of the dust, which is introduced as a random walk of the dust particles. At each time-step ${\rm d}t$, particles are displaced in a random direction by a length $l_{\rm turb} = \sqrt(D_{\rm d} {\rm d}t)$, where $D_{\rm d} = \alpha c_{\rm s} H/(1 + {\rm St}^2)$ corresponds to the diffusivity coefficient for dust \citep{Youdin+Lithwick-2007}.

Our dust simulation uses 150\,000 particles. The dust sizes $s$ range from $s_{\rm min} = 1 \mu  \rm m$ to $s_{\rm max} = 1 \rm cm$, covering the whole dust-size spectrum logarithmically. We set an initial power-law size distribution $n(s) \propto s^{-1}$, which gives a power-law dust-density distribution equal to the gas surface distribution. This choice implies approximately the same number of particles per bin size, which allows us to better follow the dust kinematics by directly comparing the assumed dust species. The dust content is introduced at the beginning of the simulation, and is assumed to be a mixture of pyroxene and ices, with an intrinsic bulk density of $\rho_d = 2.0 \rm ~ g ~ cm^{-3}$.
The initial gas-to-dust ratio is fixed to 100:1.

To calculate a physical quantity such as the effective mass accumulated in a region, or for instance the dust continuum emission, we are required to specify the size distribution $n(s)$ which can be treated as a free parameter, as well as the total dust mass. This free choice does not affect the initial dust size scaling $n(s) \propto s^{-1}$, which was chosen for computational reasons. In other words, the simulations give us the spatial distribution of particles, but we need to use a realistic size distribution to compute realistic physical quantities.

Consequently, once the simulation is complete, we re-scale the size distribution by $n(s) \propto s^{-p}$, using $p=3.5$ \citep{Dohnanyi1969}. Therefore, the dust
mass per bin size $M_i$ can be computed as
\begin{equation}
    M_i = \epsilon M_{\rm gas} \frac{s_i^{4-p}}{\sum_i s_i^{4-p}},
\end{equation}
where $\epsilon$ is the initial dust-to-gas mass ratio (i.e., 1:100), and $M_{\rm gas}$ the total mass gas. The surface density $\sigma_i$ for each $i$-bin can then be computed as
\begin{equation}
    \sigma_i = \frac{N_i(r, \phi)}{A(r, \phi)} \frac{M_i}{\sum_{i} N_i(r, \phi)},
\end{equation}
where $N_i(r, \phi)$ is the number of dust particles per bin size, $A(r, \phi)$ the surface area of each cell grid, $M_i$ the dust mass per bin size computed above.

\section{Physical considerations}\label{physical}

\subsection{Vortices}

The vortensity gives a measure of the local rotation of the fluid, and its minimum indicates the presence of a local pressure bump. Particles are attracted to higher pressure regions, even for a slight enhancement from the background \citep{Klahr+2001}. Therefore, dust accumulation is expected at the center of vortices (e.g., \citealt{Barge&Sommeria1995, Chavanis2000, Pinilla+2012, Baruteau+2019}). 

Vortices are expected to appear in the disk as a consequence of Rossby wave instabilities RWI \citep{Lovelace+1999, Li+2000} or baroclinic instabilities \cite{Klahr+2003}. Inside them, pressure maxima are present, with a minimum of the gas vortensity. Typical regions where these instabilities appear, excited by density gradient, are at the edges of the gap carved by a planet (e.g., \citealt{Lin2012}).

To visualize the vortices generated in the disk, it is useful to compute the amplitude of the vortensity perturbations relative to its initial profile, that is, $(\omega - \omega_0)/\omega_0$. The vortensity field is defined as
\begin{equation}
\boldsymbol{\omega} = \frac{\boldsymbol{\nabla} \times \boldsymbol{v}}{\rho}, 
\label{eq_vortensity}
\end{equation}
where $\textbf{v}$ is the local gas velocity, and $\rho = \Sigma / H$ is the vertically averaged density.

As our simulations are in 2D, we simply use $\omega$ to refer to the $z$-component of $\boldsymbol{\omega}$. The quantity $\omega_0$ corresponds to the value of the ($z$-component) vortensity for the initial disk profile.

\subsection{Lagrangian points $L_4$ and $L_5$} \label{Lagrangian_points_L4_L5}

\begin{figure}
\centering
\includegraphics[width=0.9\linewidth]{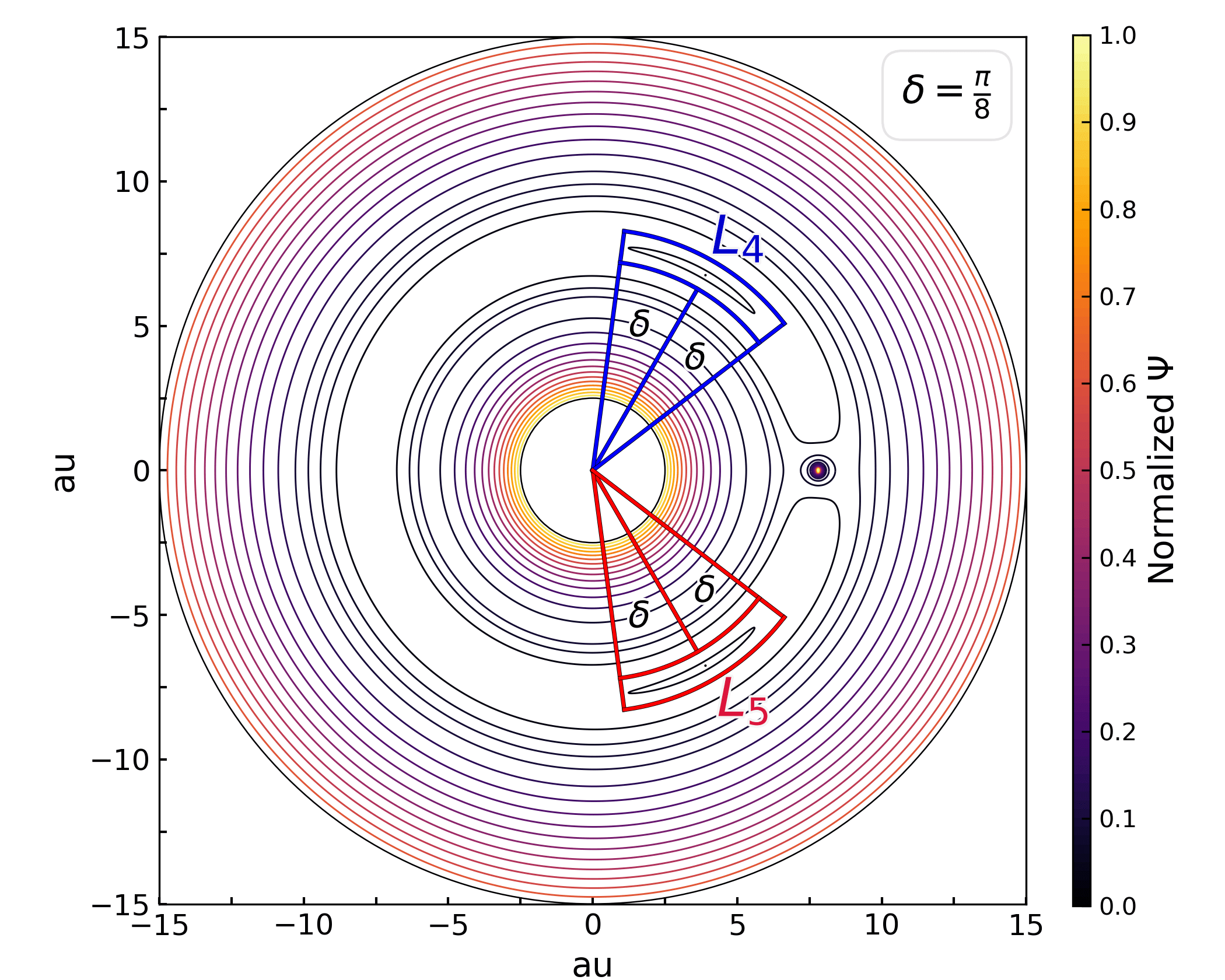}
\caption{Equipotential lines according to Eq. \ref{coPotential} in a rotating frame with the planet,
applied to a model with parameters $M_\star = 1 M_\odot$, $r_p = 7.8 \rm au$, and $M_{\rm p} = 5 M_J$. The leading Lagrangian point ($L_4$) is located at $+\pi/3$, while the trailing point ($L_5$) is at $-\pi/3$. A stable librating azimuthal angle is assumed to be $\pm \pi/8$ around each Lagrangian point. }
\label{EquipotencialLines}
\end{figure}

We are interested in the dust accumulation in the vicinity of the Lagrangian points $L_4$ and $L_5$. To compute their accretion, we need to establish the criterion of dust trapping in the vicinity of those points. A simple approximation can be obtained by solving the three-body problem consisting of two massive bodies of mass $M_\star$ (primary) and $M_{\rm p}$ (secondary), where the secondary is moving in circular orbit around their mutual center of mass, and a third particle (test particle of negligible mass) moves in the same plane. Since we are interested in the motion of the test particle, it is convenient to refer to the coordinate system of the particle. In that case, the system is rotating with an angular speed $\Omega = \sqrt{G M / a^3}$, where $M = (M_\star + M_{\rm p})$, and $a$ the separation between the two bodies.
The dynamic behavior of the third body can be obtained therefore from the effective potential
\begin{equation}
\Psi = -G \left(   \frac{M_\star}{r_1} + \frac{M_{\rm p}}{r_2}  \right) - \frac{1}{2} r^2 \Omega^2,
\label{coPotential}
\end{equation}
where $r_1$ and $r_2$ are the distances between the test particle and $M_\star$, and $M_{\rm p}$ respectively, and r is the distance to the center of mass of the two bodies (in practice the distance to $M_\star$). Equation~\eqref{coPotential} is called the  co-rotating effective potential.

In Figure \ref{EquipotencialLines} we plot the co-rotating equipotential contours from Eq. \ref{coPotential} using the parameter of our simulation: $M_\star = M_\odot$ and $M_{\rm p} = 5 M_J$, both separated by 7.8 au. The figure shows the Lagrangian points $L_4$ and $L_5$ at the planet orbital radius, corresponding to stable equilibrium positions with a zero gradient potential. $L_4$ and $L_5$ are located $\sim \pi/3$ radians in azimuth ahead of the planet and behind it, respectively.

Following the contour plot in Figure \ref{EquipotencialLines}, we consider that a particle will be in tadpole orbit around a Lagrangian point if it belongs to the cylindrical areas drawn in Figure \ref{EquipotencialLines} that surround $L_4$ and $L_5$. Each cylindrical area is delimited by
the position of the Lagrangian points at $\pm \pi/3$ adding (or subtracting) an angle of $\pm \pi/8$ rad in the azimuthal direction, and between an effective capture radius given by the range $r_{\rm p} \pm  R_{\rm Hill}$, that is,
\begin{eqnarray}
 \pi/3 ~ (L_4) - \pi/8 <  & \phi & < \pi/3 ~ (L_4) + \pi/8, \label{eqLagrangeRegion1} \\
 -\pi/3 ~ (L_5) - \pi/8 <  & \phi & < -\pi/3 ~ (L_5) + \pi/8, \label{eqLagrangeRegion2}\\
 r_{\rm p} - R_{\rm Hill}   <  & r & < r_{\rm p} + R_{\rm Hill}, \label{eqLagrangeRegion3}
 \end{eqnarray}
where $r_{\rm p}$ is the planet radial location and 
$R_{\rm Hill}$ is the Hill radius\footnote{The Hill radius is defined $R_{\rm hill} = r_{\rm p} \left( M_{\rm p} / (3 M_\star) \right )^{1/3}$, where $r_{\rm p}$ is the star--planet distance (the planet has no eccentricity in our models), $M_{\rm p}$ the mass of the planet, and $M_\star$ the mass of the star. We adopt $R_{\rm CPD} = 0.6 R_{\rm hill}$ which is a suitable choice for Jupiter-like planets around a solar mass star \citep{Crida+2009}.}. We show a posteriori that the libration of particles around a Lagrangian point occurs in such a pre-defined area.

At each time-step, we count particles that enter and leave the defined region. This method enables us to calculate the net dust flux in $L_4$ and $L_5$. At late evolutionary stages of the disk, when a stationary regime is reached, the Lagrangian points neither accumulate nor lose material.

\section{Results} \label{results}

\subsection{Gas and dust evolution}\label{GasAndDustEvolution}

\begin{figure}
\centering
\includegraphics[width=0.475\textwidth] {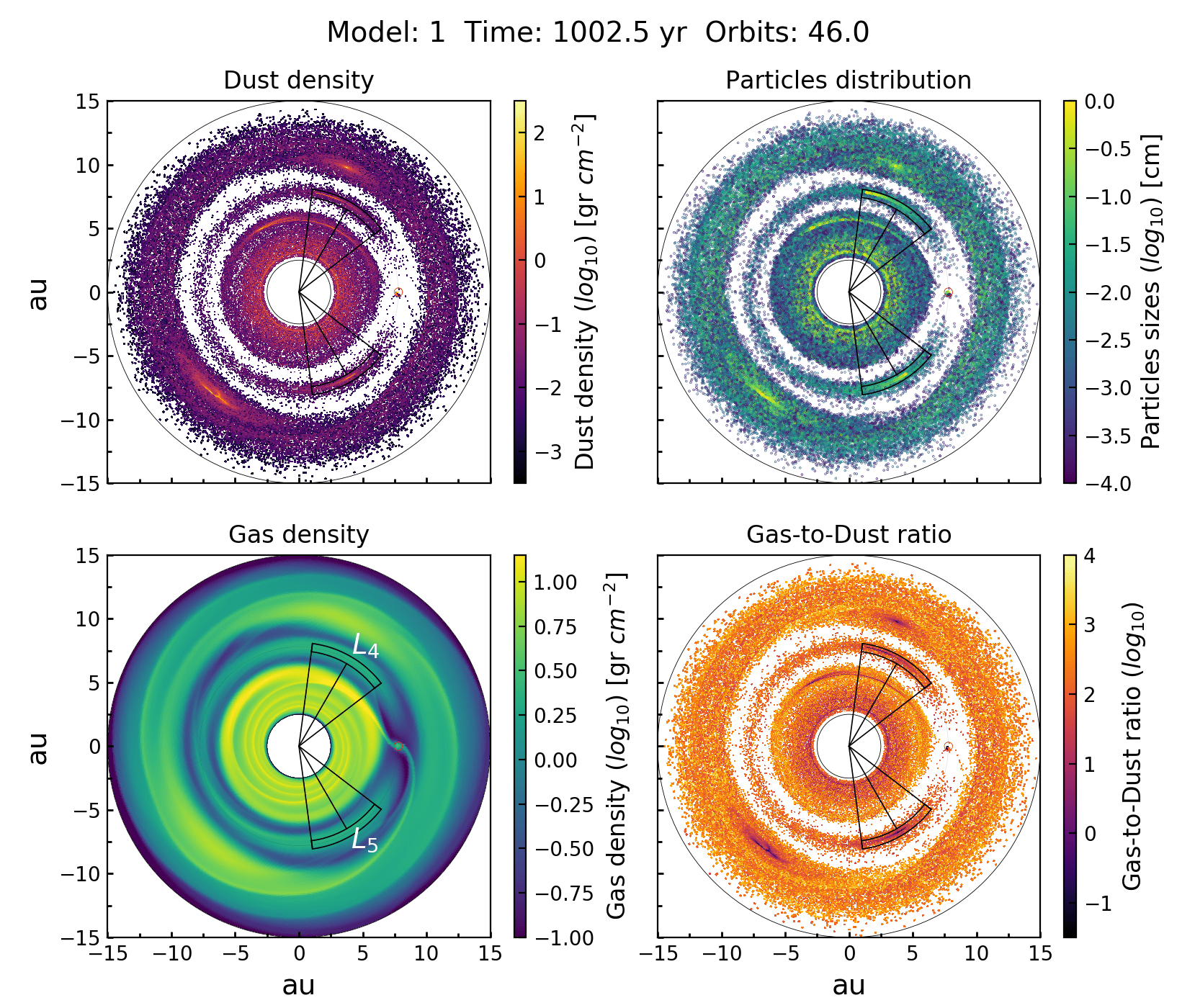}
\includegraphics[width=0.475\textwidth] {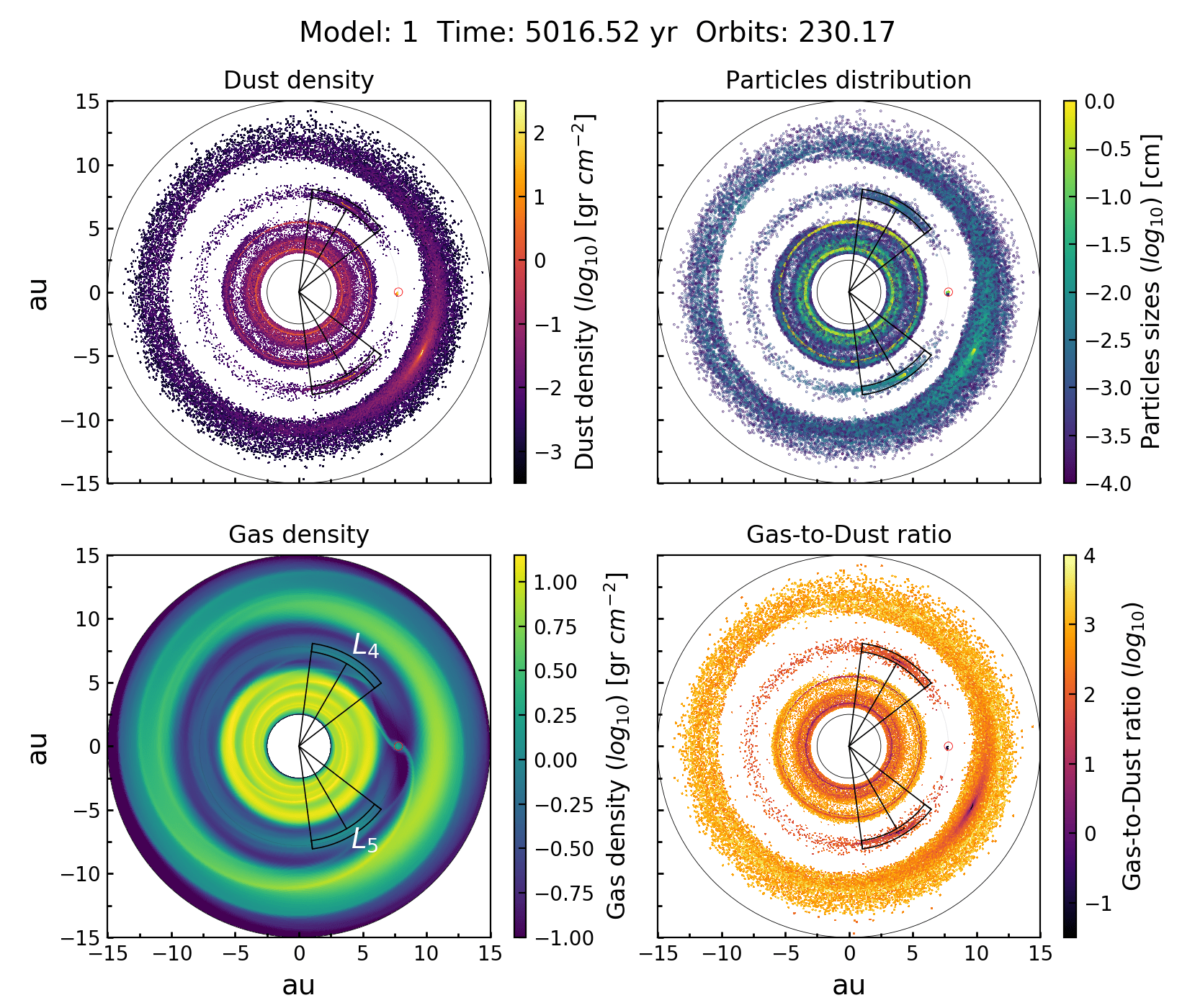}
\includegraphics[width=0.475\textwidth] {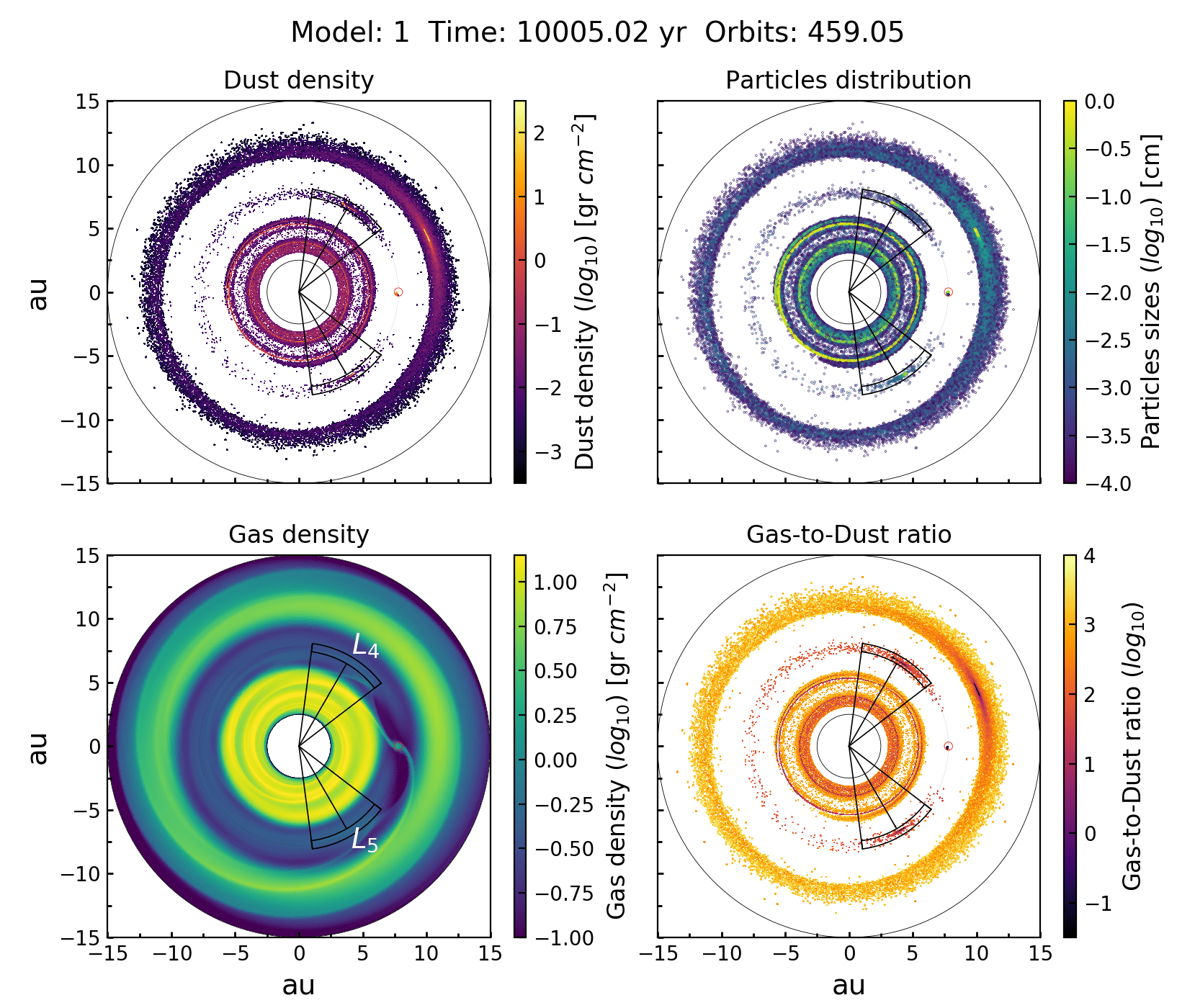}
\caption{Evolution of the gas and dust in model (1), for three evolutionary stages; 46, 230, and 459 orbits. The dust panels show the full size range of the particles, from $10^{-4}$ to $10^0$ cm. The panels have been rotated to always keep the planet at the same location x = 7.8 au, y = 0.}
\label{all_density_model1}
\end{figure}

Figure \ref{all_density_model1} shows the evolution of the disk model 1 for the timescales 46, 230, and 459 orbits (as one planetary orbit takes 21.8 yr, this corresponds to 1\,000, 5\,000, and 10\,000 yr, respectively), the different subplots include: the gas density, dust density, particle distribution, and gas-to-dust ratio for all the sizes ranging from $10^{-4}$ to $10^{0}$ cm. All models start with a gas-to-dust ratio equal to 100:1.

During the evolution, we observe the formation of an inner disk and an outer one as the planetary gap develops, along with typical density wakes at the Lindblad resonances. By the end of the simulation, the planet practically depletes  the co-rotating zone from gas (bottom of Figure \ref{all_density_model1}).
As expected, the dust evolves differently depending on its coupling to the gas.

At 46 orbits (upper-panel Figure \ref{all_density_model1}), a gap is already present in the disk. However, a horseshoe structure of dust with particles of all sizes populates the co-rotating zone. The dust density inside the gap is very low $\sim 0.003 \rm~  gr ~ cm^{-2}$, with a gas-to-dust ratio of $\sim 528:1$, except at the two Lagrangian points, where the gas-to-dust ratio is reduced to 13:1 ($L_4$) and 9:1 ($L_5$), indicating dust accumulation. In the outer disk, two blobs of dust appear, corresponding to two vortices (see Sect. \ref{vortices}). The vortices are not noticeable in the gas. The gas-to-dust ratio at the large swarm reach $\sim 8:1$  (Figure \ref{all_density_model1}: coordinates $x \sim -7; y \sim -8 \rm ~ au$), 
showing efficient dust trapping.

After 230 orbits (middle panel Figure \ref{all_density_model1}), the gap is more depleted of gas, while the co-rotation zone is still populated by dust of all sizes. The mm-cm particles inside this region continue to accumulate around the Lagrangian points. The gas-to-dust ratio in these points is reduced to 4.4:1  ($L_4$) and 3.8:1  ($L_5$). The two dusty blobs present in the outer disk at 46 orbits have collided to form a single one (located at coordinates $x \sim 10; y \sim -5$ au), with a total mass of 16.3 $M_{\rm moon}$, and a gas-to-dust ratio of about 13.8:1.

At the end of the simulation (orbit 459, bottom panel Figure \ref{all_density_model1}), $\mu$m and submm particles are librating in horseshoe orbits around $L_4$ and $L_5$, while mm-cm particles are confined in tadpole orbits around either $L_4$ or $L_5$. From Figure \ref{all_density_model1} (orbit 459), the leading swarm ($L_4$) accumulate more small particles than the trailing Lagrangian point ($L_5$). However, there is more mass accumulated in $L_5$ than in $L_4$ since most of the mass is concentrated in large particles (mm-cm), which are concentrated mostly around $L_5$.

\subsection{Vortices}\label{vortices}

\begin{figure*}
\centering
\includegraphics[width=0.9\linewidth]{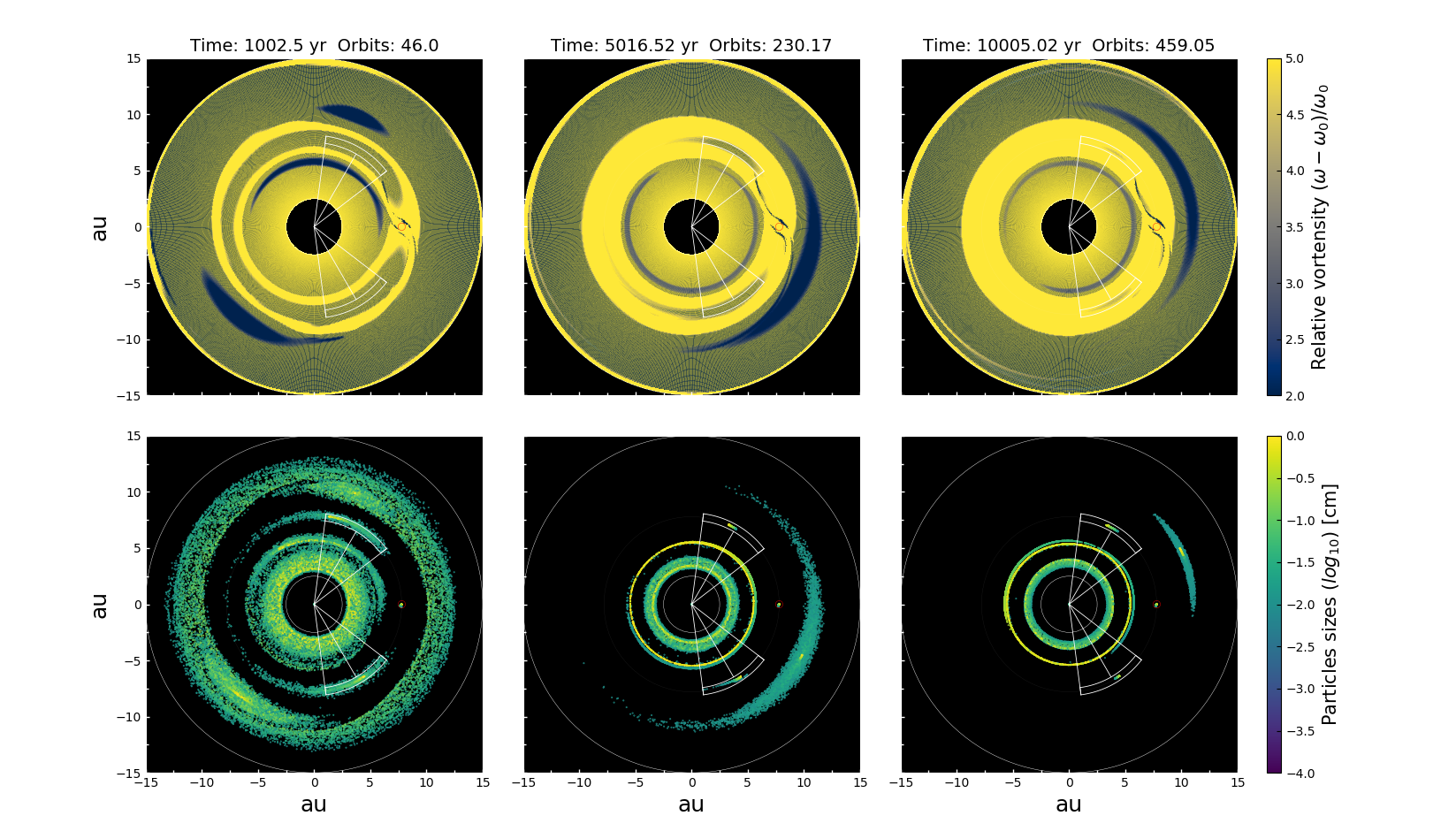}
\caption{\textit{Top}: Evolution of the vortensity and dust in model 1 , for three evolutionary stages: 46, 230, and 459 orbits. \textit{Bottom}: Dust distribution of particles with sizes in the range $10^{-2} - 10^0$
cm. The planet rotates in the counter-clockwise direction. The panels have been rotated to always keep the planet at the same location, namely $x = 7.8$ au, $y = 0$.}
\label{vortensity_10mil}
\end{figure*}

Figure \ref{vortensity_10mil} shows the amplitude of the vortensity deviations from the background for model 1, that is, $(\omega - \omega_0)/\omega_0$,  where Eq. \ref{eq_vortensity} was used to compute $\omega$ for the outputs: 46, 230, 460 orbits. For comparison, we plot  the dust distribution by sizes next to the vortensity. For a better analysis, we divide the disk into three regions: (a) the inner disk, (b) the gap, and (c) the outer disk, as described below.

(a) 
In early stages, the inner disk shows a homogeneous vortensity, with a local minimum (blue color) developing at the edge of the inner disk just next to the gap. The bottom panel of Fig. 3 shows the size distribution of dust, where sub-millimeter to cm particles (yellow color) are observed. As the disk evolves, two dusty rings concentrating submm to cm grains develop. The location of such rings coincides with the location of the minimum in vortensity.

(b) The gap exhibits a local minimum vortensity with a typical horseshoe structure, trapping submm and cm size particles. At early stages, two prominent vortices are identified at the Lagrangian points $L_4$ and $L_5$, where $L_5$ covers a larger area than $L_4$ (two blue islands observed in the gap). Hence, the Lagrangian points start to collect large particles (mm-cm). The vortices persist during the simulation, trapping more dust particles as the disk evolves. Trojan dust of $10^{-2} - 10^0$ cm lies in $L_4$ and $L_5$ at the end of the simulation.

(c)  Initially, the external disk exhibits two notorious vortices (blue islands), which start to trap dust. With time, the two vortices collide to form one single vortex, resulting in a larger banana-shaped vortex, concentrating submm to cm particles. The dust mass of the final outer lobe or satellite reaches $\sim 16.6$ $M_{\rm moon}$ with a gas-to-dust ratio of about 9:1. One should note that the dust distribution in Figure \ref{vortensity_10mil} only shows submm to cm grains because these sizes are effectively trapped in the local pressure maximum and vortices.

At the end of the simulation, mm to cm particles have been efficiently trapped at the Lagrangian points by small vortices created by the planet--star system. The vortices continue to be present at the end of the simulation although the size of the now-single vortex has reduced over time. If there were more material within the gap, the slightly higher  pressure at the center of the vortices would continue to trap dust.
Besides the populated Lagrangian points, two concentric rings at the inner region develop. These are dust traps created by planetary wakes.

\subsection{Dust accumulation by sizes around Lagrangian points}\label{DustLagrangianSection}

\subsubsection{Gas--dust interaction}

\begin{figure*}
\centering
\includegraphics[width=1.\linewidth]{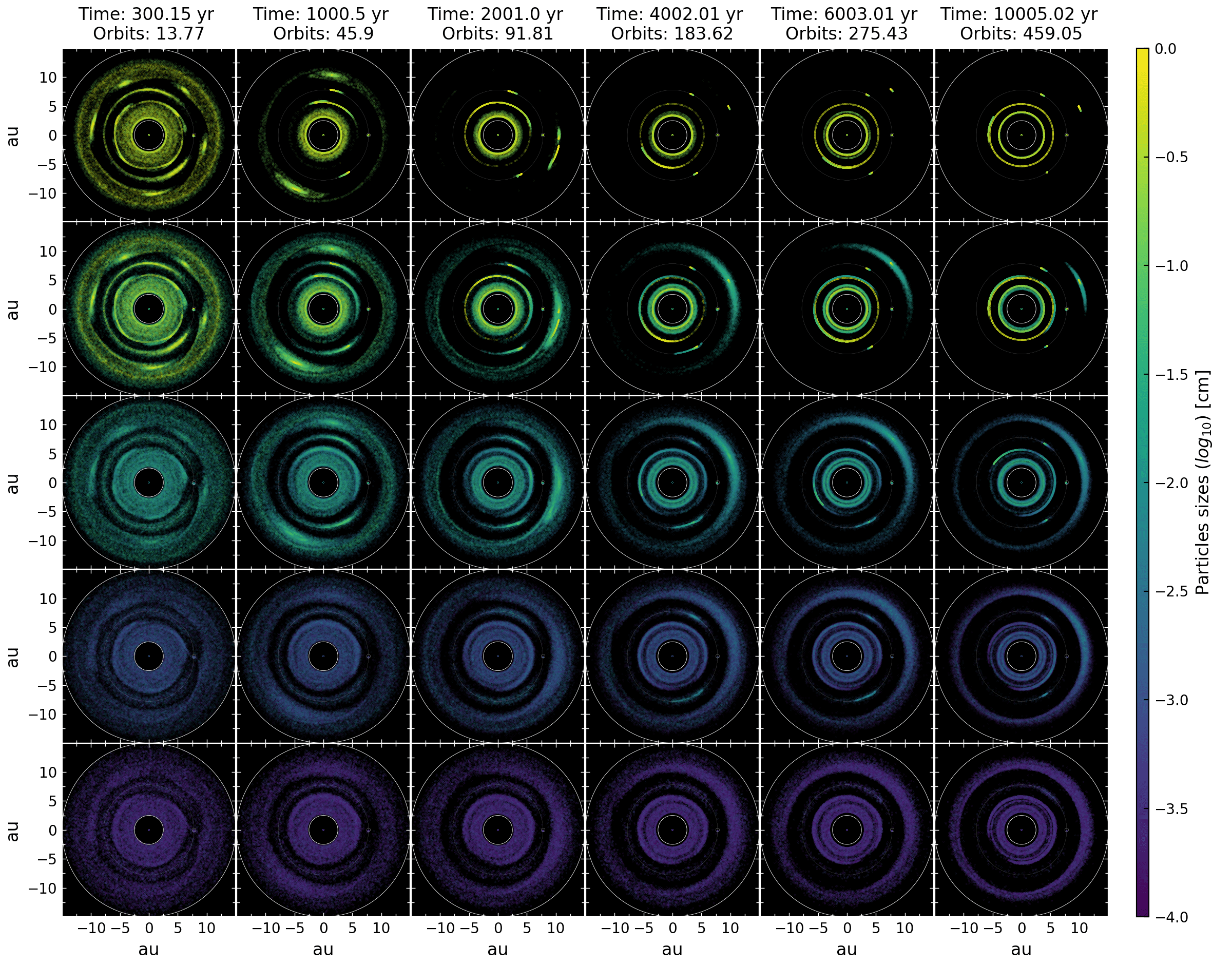}
\caption{Evolution in time of the dust for model 1 as a function of particle size, ranging from $\mu$m to cm. The panels have been rotated to always keep the planet at the same location x = 7.8 au, y = 0.}
\label{dust_evolucion_model1}
\end{figure*}

Figure~\ref{dust_evolucion_model1} shows the evolution for each dust bin size ($\mu$m to cm) 
at different evolutionary stages: 300, 2\,000, 4\,000,  6\,000, and 10\,000~yr. We observe that the largest particles in the range of 0.3-1 cm (yellow) quickly accumulate at the Lagrangian points. In less than 1\,000 yr (45 orbits), $L_4$ and $L_5$ are already settled, including a forming system of two inner rings (at $\sim 4$ au), originated from planetary wakes. On the same timescale, small particles in the range of $10^{-1.5} - 10^{-1}$ cm (green color) are also trapped at the Lagrangian points. At the end of the simulation, a relatively massive swarm of dust of $\sim$ 16.6 $M_{\rm moon}$ is formed  by mm-cm particles (yellow/green) located at the outer region showing a banana-shaped dust concentration. The swarm corresponds to the leftover of the extinct external dusty disk.

Particle sizes in the range of $10^{-2.5}$ - $10^{-1.5}$ cm (green-blue) behave in a slightly different fashion. A horseshoe starts to develop at 1\,000 yr. As the disk evolves, a gap is carved by the planet. Soon, the gap reduces its number of green-blue particles, which remain mostly trapped at the Lagrangian points. An external disk is also created as the dust evolves. The external swarm of 16.6 $M_{\rm moon}$ described above is hidden in this outer ring of $10^{-2.5}$ - $10^{-1.5}$ cm particles. Micrometer to $10^{-2.5}$ cm (blue-violet) particles are distributed all over the disk; in the inner disk, within the gap (librating in horseshoe orbits around $L_4$ and $L_5$), and in  the outer disk.

\begin{figure}
\centering
\includegraphics[width=0.9\linewidth]{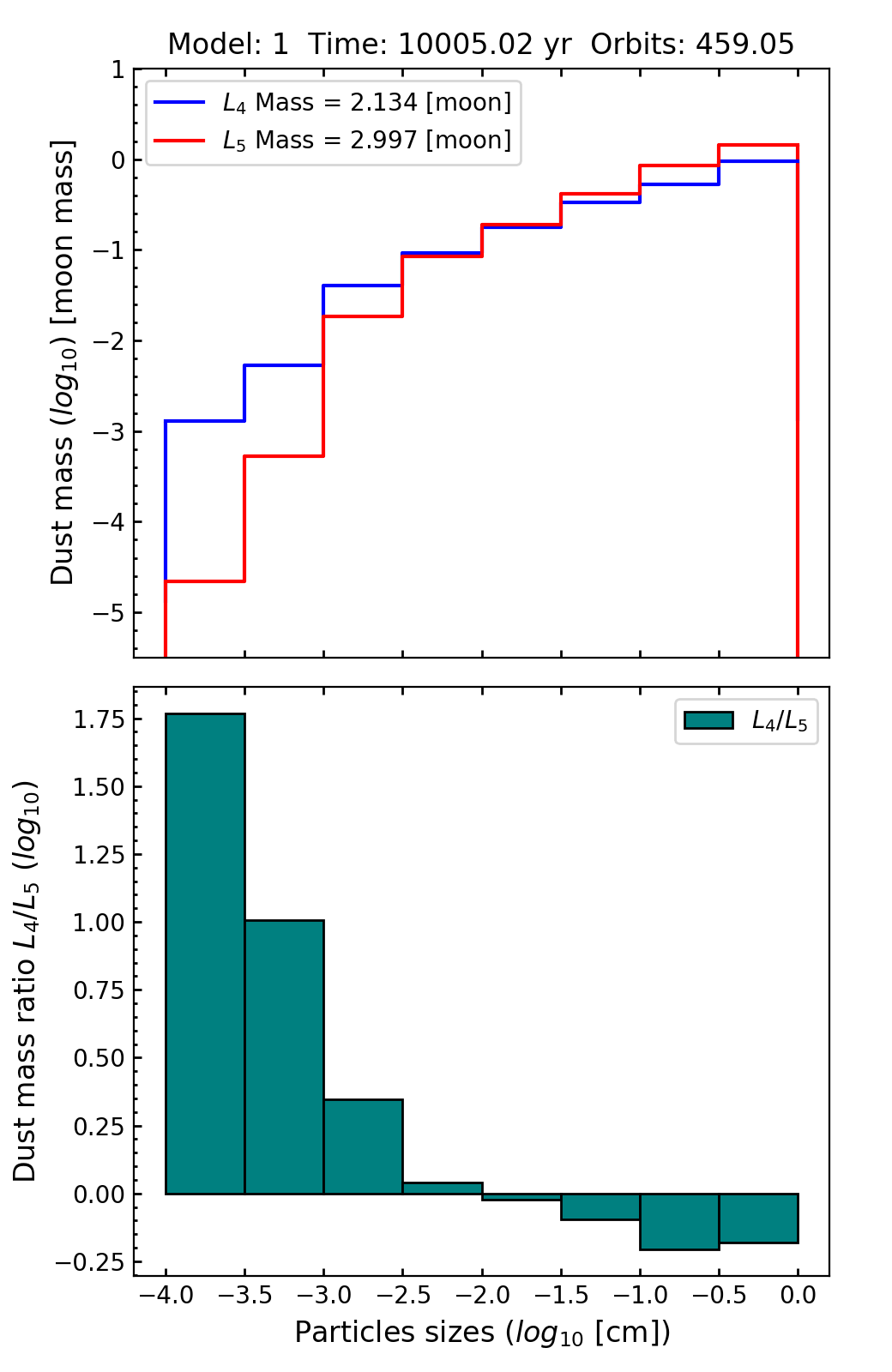}
\caption{Mass spectrum of model 1 as a function of the size of the dust accumulated in $L_4$ and $L_5$ for the last evolutionary stage (460 orbits). The bottom panel shows the mass ratio $L_4/L_5$ as a function of dust particle size.}
\label{histo_L4_L5_model1}
\end{figure}

\begin{figure}
\centering
\includegraphics[width=0.9\linewidth]{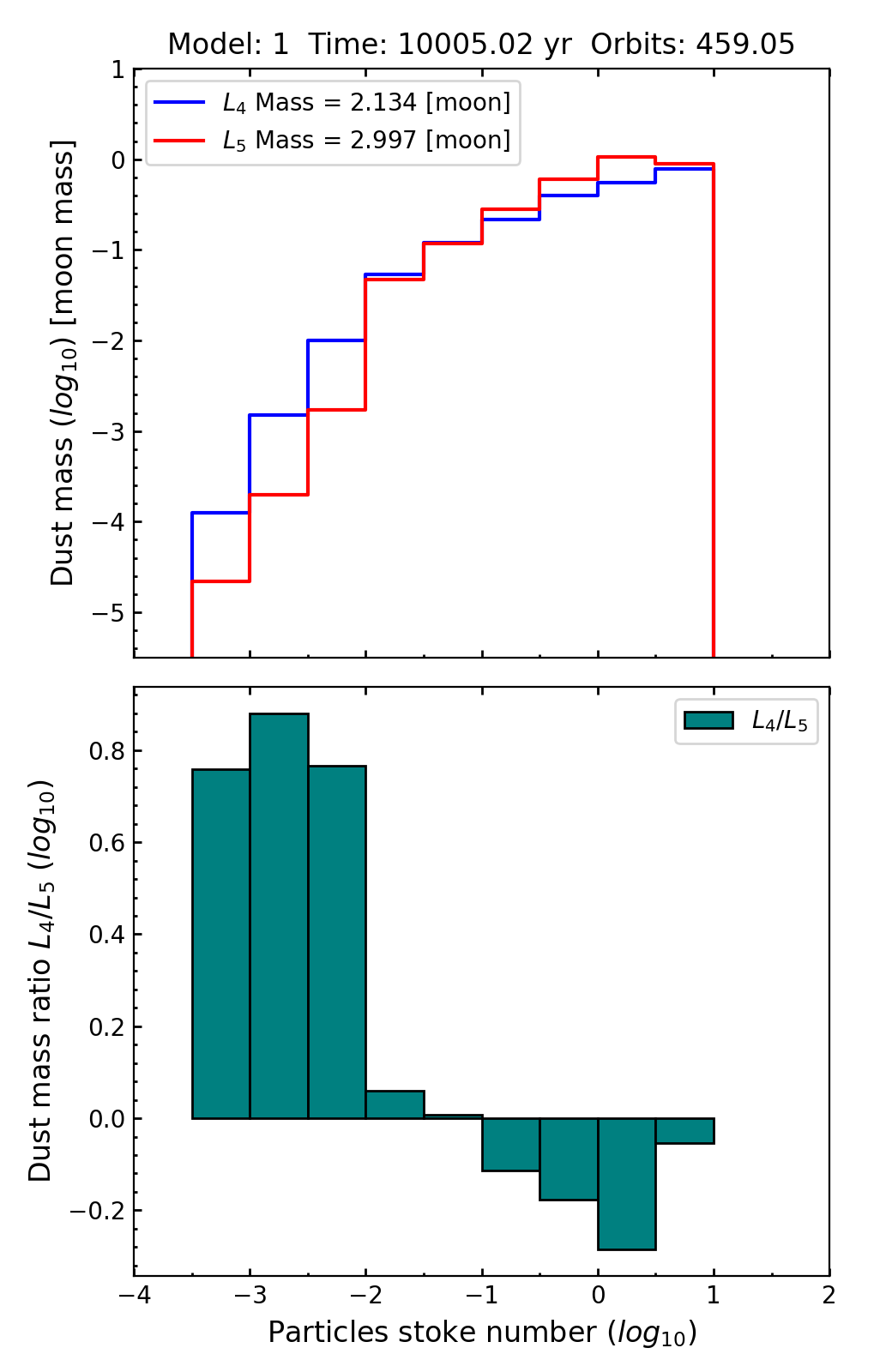}
\caption{Mass spectrum of model 1 as a function of the Stokes number for the dust accumulated in $L_4$ and $L_5$ at the last evolutionary stage (460 orbits). The bottom panel shows the mass ratio $L_4/L_5$.}
\label{histo_L4_L5_Stokes_model1}
\end{figure}

Once the simulation ends, we analyze the distribution of particle size and Stokes number around the Lagrangian points, computing how much material is finally concentrated in $L_4$ and $L_5$. In Figure \ref{histo_L4_L5_model1} and \ref{histo_L4_L5_Stokes_model1} we plot  the mass spectrum of $L_4$ and $L_5$ for model 1, and their ratio $L_4/L_5$ after 10\,000 yr (460 orbits) of evolution as a function of particle size and Stokes number, respectively.

\begin{figure}
\centering
\includegraphics[width=0.9\linewidth]{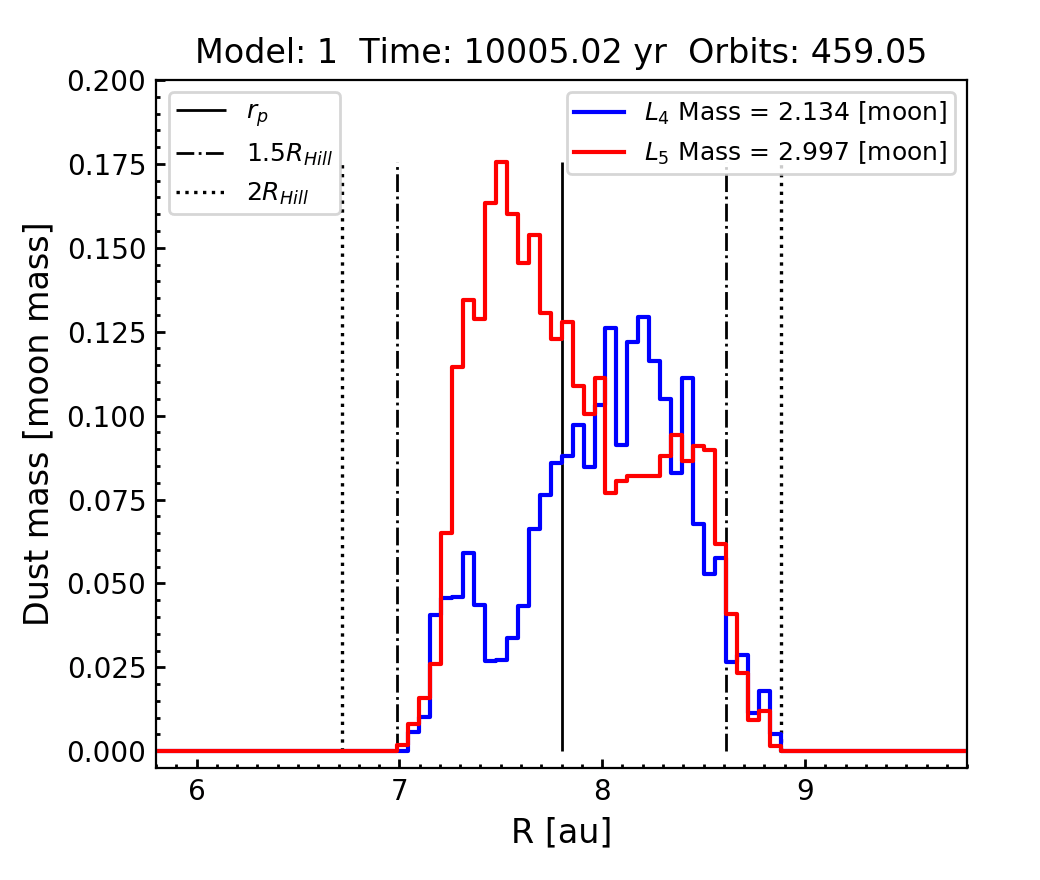}
\caption{Histogram of the effective dust mass accumulated at the end of the simulation inside $L_4$ and $L_5$, distributed by radial bins. This defines a capture region given by $r_{\rm p} + 2 R_{\rm Hill}$ from the planet towards the outer region, and $r_{\rm p} - 1.5 R_{\rm Hill}$ from the planet towards the star. An \textit{effective} capture radius is expected at $r_{\rm p} \pm R_{\rm hill}$.}
\label{Histograma_Initial_random}
\end{figure}

In Figure \ref{Histograma_Initial_random}, we plot the accumulated effective mass per radial bin. We notice that centering an annulus at the planet location $r_{\rm p}$, the capture area extends from $r_{\rm p}+ 2 R_{\rm Hill}$ towards the outer region, and $r_{\rm p} - 1.5 R_{\rm Hill}$ towards the star. However, the \textit{effective} capture radius, defining tadpole orbits, is mostly inside $r_{\rm p} \pm R_{\rm Hill}$ as defined by Eq. \ref{eqLagrangeRegion3}.

We observe the following interesting features from the mass spectrum: (i) There is an asymmetry in the total mass accumulated in each Lagrangian point. $L_4$ accumulates 2.1 $M_{\rm moon}$, while $L_5$ accumulates 3.0 $M_{\rm moon}$ (Figs. \ref{histo_L4_L5_model1} and \ref{Histograma_Initial_random}). (ii) Most of the effective mass trapped in $L_4$ and $L_5$ is foumd in particles in the range of $\sim 0.03$ to $1$ cm (top panel of Fig. \ref{histo_L4_L5_model1}). A small amount of micron-sized particles are also trapped in the Lagrangian points. (iii) The mass asymmetry between $L_4$ and $L_5$ depends on the particle size. Small particles in the range $10^{-4}$ to $10^{-1.5}$ cm are more abundant in $L_4$, while particle of  $10^{-1.5}$ to 1 cm are more abundant in $L_5$ (see bottom panel of Figure \ref{histo_L4_L5_model1}). The transition (when mass in $L_4$ is equal to that in $L_5$) happens for $10^{-2}$ cm particles. However, the effective mass (all sizes) is always larger in $L_5$ than $L_4$. (iv) The range of the Stokes number of particles trapped in the Lagrangian points lies in the range ${\rm St} \sim 10^{-6} - 10$. Particles trapped in $L_4$ (small particles) are dominated by Stokes number ${\rm St} \textless 0.1$.  Particles with large Stokes numbers, that is with ${\rm St} \textgreater 0.1,$ are commonly trapped in $L_5$.

\subsubsection{Mass accretion at Lagrangian points}

\begin{figure*}
\centering
\includegraphics[width=0.9\linewidth]{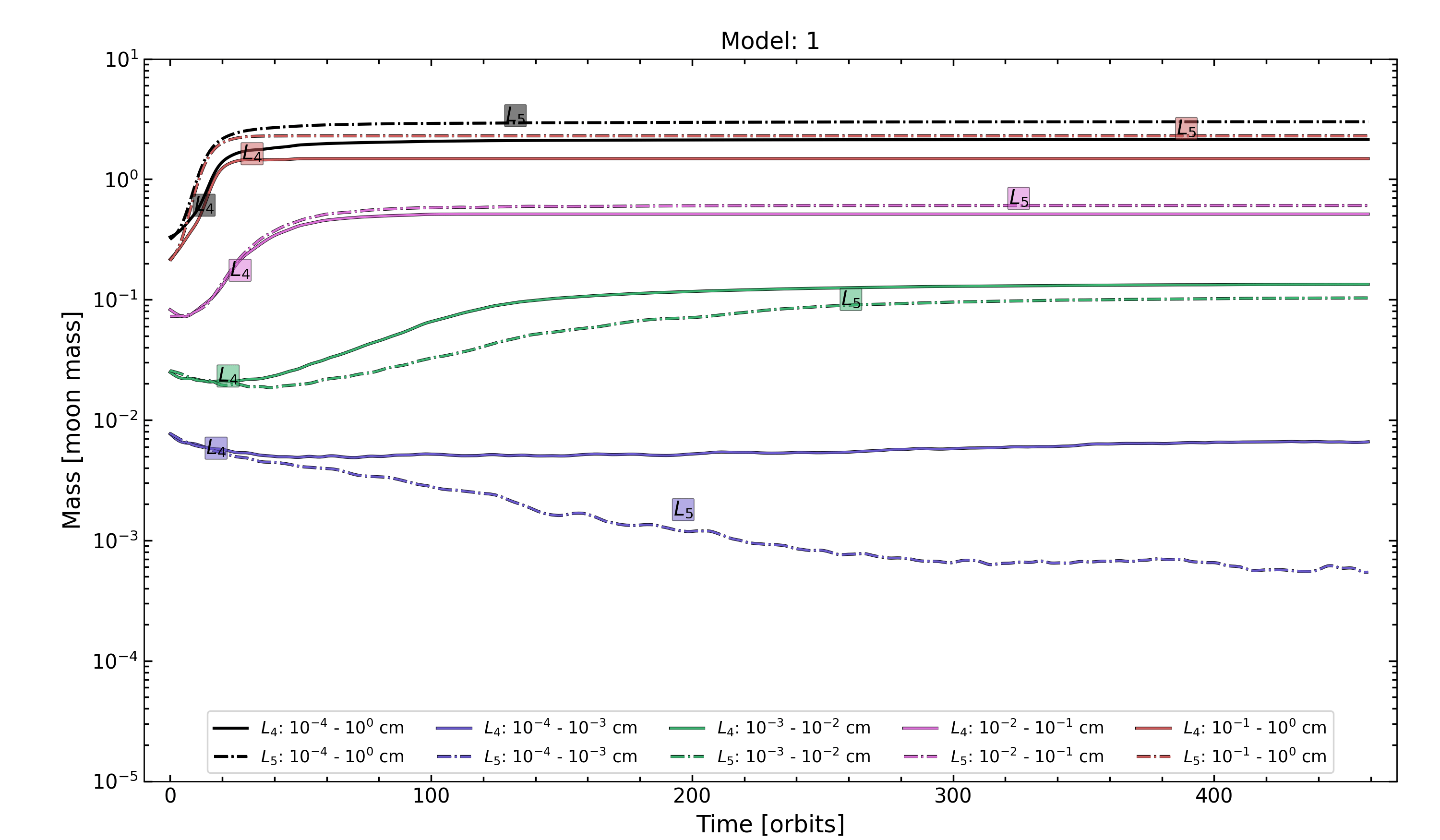}
\caption{Dust mass accumulated in time around the Lagrangian points $L_4$ (continuous line) and $L_5$ (dotted line) for model 1. The evolution is shown
for different sizes, including the effective (all sizes) accumulated mass (black continuous line).}
\label{L4_L5_mass_evolution_model1}
\end{figure*}

The dust accumulation rate around a Lagrangian point differs depending on the specific particle size. 
Figure \ref{L4_L5_mass_evolution_model1} shows mass as a function of time in the vicinity of Lagrangian points for a specific range of sizes for model 1, that is: $10^{-4}-10^{0}$, $10^{-4}-10^{-3}$, $10^{-3}-10^{-2}$, $10^{-2}-10^{-1}$, and $10^{-1}-10^{0}$ cm. Initially, the total dust mass grows exponentially. During the first 50 planetary orbits, $L_5$ accumulates  $\sim 2.7$ $M_{\rm moon}$  of dust, while $L_4$ reaches about $\sim 1.9$ $M_{\rm moon}$, where 
the main contribution of the mass comes from the largest particle size, namely, $10^{-1} - 10^{0}$ cm. After that, the accretion rate is almost completely halted, reaching its final mass of 2.13 $M_{\rm moon}$ for $L_4$, and 3.0 $M_{\rm moon}$ for $L_5$.

Lagrangian points are almost completely depleted of micron-size particles. Figure \ref{L4_L5_mass_evolution_model1} shows that particles in the range of $10^{-4} - 10^{-3}$ cm do not contribute to the mass trapped in $L_4$ and $L_5$. The generated vortex at the Lagrangian points is not strong enough to attach stable orbits for this size  range. However, a considerable amount of gas is located around $L_4$ and $L_5$, namely 4.8 $M_{\rm moon}$ of gas in $L_4$, and 5.6 $M_{\rm moon}$ in $L_5$ (with mean gas density of $\sim 0.4$ and $ \sim 0.5 ~ \rm g~cm^{-2}$ respectively), which is responsible for the pressure bump trapping mm-cm dust. At the end of the simulation, the gas-to-dust ratio is 2.2:1  for $L_4$ and 1.9:1 for $L_5$. 

Table \ref{TableLpoints}  summarizes the dust properties at the Lagrangian points at the end of each simulation, for all the models in Table \ref{models_parameters}. $M_{\rm {L_4,L_5}}$ represents the accumulated dust mass, $\sigma_{\rm {L_4,L_5}}^{\rm dust}$ the dust density, $\Sigma_{\rm gas}/\sigma_{\rm dust}$ the gas-to-dust ratio, and $<T>$ the mean dust temperature, where we assume that the dust has the same temperature as the gas. All these quantities are evaluated at each Lagrangian point $L_4$ and $L_5$, inside the effective capture region given by Equations
~\eqref{eqLagrangeRegion1}-\eqref{eqLagrangeRegion3}.

\begin{table*}
\begin{center}
\begin{tabular}{ccccccccc}
    \hline
    \hline
    \multicolumn{7}{c}{  $L_4$ and $L_5$ dust properties}\\
    \hline
    \hline
    \textbf{Model} & $M_{L_4}$ [moon] & $M_{L_5}$ [moon] & $\sigma_{\rm L_4}^{\rm dust}[g \, cm^{-2}]$ & $\sigma_{\rm L_5}^{\rm dust}[g \, cm^{-2}]$  & $(\Sigma_{\rm gas}/\sigma_{\rm dust})_{\rm L_4}$ & $(\Sigma_{\rm gas}/\sigma_{\rm dust})_{\rm L_5}$ & $<T_{L_4}>$ [K] & $<T_{L_5}>$ [K]\\ 
     1 &  2.13   &3.0 &0.18  &0.25  &2.23 &1.9 & 57  & 55\\
     2 &  0.60   &2.23 &0.049  &0.184  &5.29 &1.48 & 63 & 63\\
     3 &   0.001 &2.8    & $4.49 \times 10^{-5}$&0.14 &5808.6& 3.31 & 66 & 58 \\
     4 &  0.06   &2.4    &0.003 &0.12  &71.9    &4.85  & 44 & 33 \\
     5 &  1.12   &2.91 &0.054  &0.140  &5.50 &2.06 & 38 & 39\\
     6 &  0.88   &0.86 &0.073  &0.071  &3.37 &3.56 &64 & 64\\
    \hline
\end{tabular}
\end{center}
\caption{Dust properties at the Lagrangian points $L_4$ and $L_5$ computed for the last evolutionary stage (460 orbits). $M_{\rm {L_4,L_5}}$ represents the accumulated dust mass, $\sigma_{\rm {L_4,L_5}}^{\rm dust}$ the dust density, $\Sigma_{\rm gas}/\sigma_{\rm dust}$ the gas-to-dust ratio, $<T>$ the mean dust temperature, evaluated inside the capture region Equations \ref{eqLagrangeRegion1}-\ref{eqLagrangeRegion3}.}
\label{TableLpoints}
\end{table*}

\subsection{Instabilities at the Lagrangian points}

Our fiducial model (model 1) shows a relatively stable behavior of mass accumulation at the Lagrangian points (Figure \ref{L4_L5_mass_evolution_model1}). 
We explore the impact of turbulent viscosity $\alpha$, planetary mass $M_{\rm p}$, and stellar irradiation $L_\star$ on the stability of trapped dust. We present our findings below.

\subsubsection{Effect of turbulent viscosity}

%comparar modelos 1, 2, 6. (4 and 5)

The situation is somewhat different if the turbulent viscosity is increased. Models 1, 2, and 6 share the same parameters, apart from the viscosity: $\alpha = 10^{-4}$ (model1), $\alpha = 10^{-2}$ (model2), and $\alpha = 10^{-3}$ (model6) (see Table \ref{models_parameters}).

In Figure \ref{L4_L5_mass_evolution_model2_new}, we plot the dust-mass evolution around $L_4$ (solid line) and $L_5$ (dotted line) for model 2. In this enhanced viscosity model ($\alpha = 10^{-2}$), the effective mass reaches final values of 0.6 ($L_4$) and 2.2 ($L_5$) $M_{\rm moon}$. This is in stark contrast to the accumulated mass in the low viscous model 1 ($\alpha = 10^{-4}$), where the final mass reaches 2.1 ($L_4$) and 3.0 ($L_5$)  $M_{\rm moon}$.

Another remarkable difference from model 1 is that the only particles that effectively accumulate in both Lagrangian points are in the range of 0.3-1 cm. Small particles in the range of $10^{-2}-10^{-1}$ cm accumulate in $L_4$ and  $L_5$ for a short period only. An instability is triggered after 200 orbits, evacuating these particles from the Lagrangian points (Figure \ref{L4_L5_mass_evolution_model2_new}). When the gap is being carved, small particles initially located in co-rotating orbits are dragged towards the inner disk by gas accretion. They never follow stable tadpole orbits around the Lagrangian points for such a high-viscosity  model. Furthermore,  increasing the gas viscosity increases the transport of angular momentum, promoting gas accretion towards the star. The enhanced accretion produces a less active vortensity. The less active vortex does not retain the smaller ($10^{-4} - 10^{-2}$ cm)  particles efficiently. 

There is also the effect of diffusion of dust particles. In our simulations, such diffusion is modeled through the coefficient $D_{\rm d} = \alpha c_{\rm s} H/(1 + {\rm St}^2)$, which is proportional to the turbulent $\alpha$-viscosity. The enhanced viscosity promotes radial diffusion of particles, helping to trigger the observed instabilities, and making the accumulation of dust particles more difficult in the Lagrangian points. However, despite having the effect of higher turbulent viscosity and a diffusive mechanism for particles, we still have a significant accumulation of mm-cm particles in $L_5$.

Figure \ref{histo_L4_L5_model2} (top panel) shows the mass spectrum of the dust accumulated in the Lagrangian points for model 2. The bottom panel shows the mass ratio $L_4/L_5$ per bin size. The total mass ratio gives 0.3 (compared to 0.7 for model 1). This comparison suggests that a higher turbulence in the disk stimulates the evacuation of dust grains from $L_4$ rather than $L_5$ (therefore the mass ratio $L_4/L_5$ is reduced for high-viscosity models). In other words, the lower the viscosity, the greater the similarity between $L_4$ and $L_5$ in terms of dust
mass.

With an intermediate value for the turbulent viscosity ($\alpha = 10^{-3}$, model 6), we find an intermediate situation between model 1 ($\alpha = 10^{-4}$) and model 2 ($\alpha = 10^{-2}$). 
See Figure \ref{L4_L5_mass_evolution_model6_new} for the dust accumulation rate, and Figure \ref{histo_L4_L5_model6} for the final mass spectrum of model 6.

\subsubsection{Effect of planetary mass}

In model 3, the mass of the planet is increased to $5 M_J$, keeping the same parameter space as in model 1. The massive planet of model 3 carves a broader gap by a factor $M_{\rm p}^{1/2} = 5^{1/2}$ \citep{Kanagawa+2016}, and on a shorter timescale compared to model 1. Gravitational perturbations from the planet severely affect the evolution of dust particles around the Lagrangian points.

Figure \ref{L4_L5_mass_evolution_model3_new} shows the dust evolution around $L_4$ (solid lines) and $L_5$ (dotted lines) for model 3. In this case, only $L_5$ accumulates grains, where most of them are large particles in the range $10^{-1} - 10^{0} \rm cm$. 

The rapid formation of the gap for this model leads to an enhanced evacuation of gas inside the co-rotation zone, through a region slightly closer to the leading zone of the planet, thereby removing most of the particles attached to $L_4$. The larger the planetary mass, the less efficient the dust capture at a Lagrangian point.

\subsubsection{Effect of stellar irradiation}

Comparing models 3 and 4 reveals some aspects of the influence of stellar irradiation. Model 3 has a central star with $5 L_\odot$, while model 4 is a colder model with $1 L_\odot$ . In Model 3 (described in Sect. 4.4.2), particles at $L_4$ are completely decoupled after 200 orbits of evolution. 
For a colder model such as model 4, particles that once accumulated around $L_4$ also become unbound as a consequence of the huge gap created by the massive $5 M_J $planet), but at a later evolutionary stage, after 340 orbits (compared to 200 orbits for the hotter model 3). The broader gap leads to particles decoupling from $L_4$ anyway.
In a colder disk, the pressure scale-height is smaller, making it easier for a planet to open a gap and produce variations in the pressure field with sharper transitions in the radial direction, which translates to the promotion of RWI and hence the formation of vortices that last longer.

\subsection{Evolution and trajectory of dust particles}\label{EvolDustTraj}

For better visualization, we divide the last evolutionary stage of model 1 into three regions: (a) inner ring, (b) the gap, and (c) the outer disk. Each zone features different morphologies worth studying (see Figure \ref{dust_evolucion_model1}).  We randomly select a number of particles of different sizes from the last step of the simulation belonging to specific zones of the above three regions and trace back along their evolutionary path, starting from the initial time-step to the last one.

\subsubsection{Inner ring}

\begin{figure}
\centering
\includegraphics[width=0.9\linewidth]{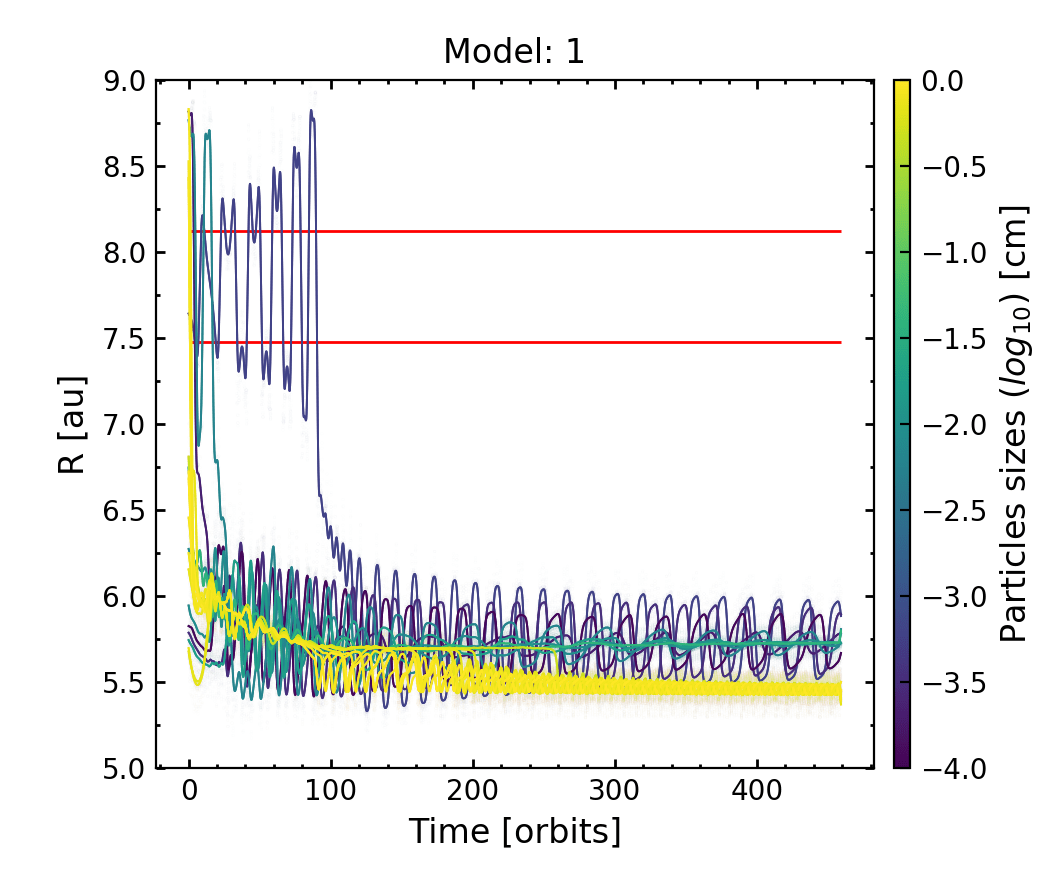}
\caption{Radial trajectory of randomly selected particles ($\mu$m to cm) that end up trapped in the second inner ring located at $r = 5.6$ au featured at the end of the simulation in model 1.}
\label{RvsTime_inner_ring_model1}
\end{figure}

At the inner rim of the disk, two concentric mm-cm dusty rings produced by planetary wakes are located at $r \sim 3.8$, and $r \sim 5.6$ au, revolving at resonances of 3:1 (3.8 au) and 5:3 (5.6 au), respectively, with respect to the planet. The rings are clearly visible at the end of the simulation (top right panel of Figure \ref{dust_evolucion_model1}). 

We selected particles belonging to the second dusty ring at $r = 5.6$ au (for the output 459 orbits, model 1) and traced back their trajectories. Figure \ref{RvsTime_inner_ring_model1} shows the evolution of their radial position as a function of time, starting at $t = 0$ to $t = 10\,000$ yr. The color bar indicates the size of each particle. The two horizontal red lines indicate the position of the Lagrangian libration zone (the planet is located at $r_{\rm p}= 7.8 \rm au$). The particles that end up  trapped at the inner ring ($r = 5.6 \rm au$) came from different regions of the disk, initially located between the radii at 5.5au and 8.8 au (see $t=0$ in Figure \ref{RvsTime_inner_ring_model1}).

The mm-cm particles beyond 6.5 au migrate due to drag forces to their final position at 5.6 au (Figure \ref{RvsTime_inner_ring_model1}), while $\mu$m particles of this ring were initially located at the same original radial distance (close to 5.6 au), oscillating with an amplitude of $ \sim  1$ au around $r \sim 5.6$ au while they travel through the full orbital path around the star. A few $\mu$m particles were initially found at the co-rotation zone of the planet;  being coupled to the gas, these particles were relocated to the inner ring by gas accretion.
It is interesting to note that the ring located at $r \sim 5.6$ au is rather composed of two close small rings; one accumulates submm particles, while the other accumulates mm-cm particles.

\subsubsection{The gap: tadpole and horseshoe orbits}

\begin{figure}
\centering
\includegraphics[width=0.9\linewidth]{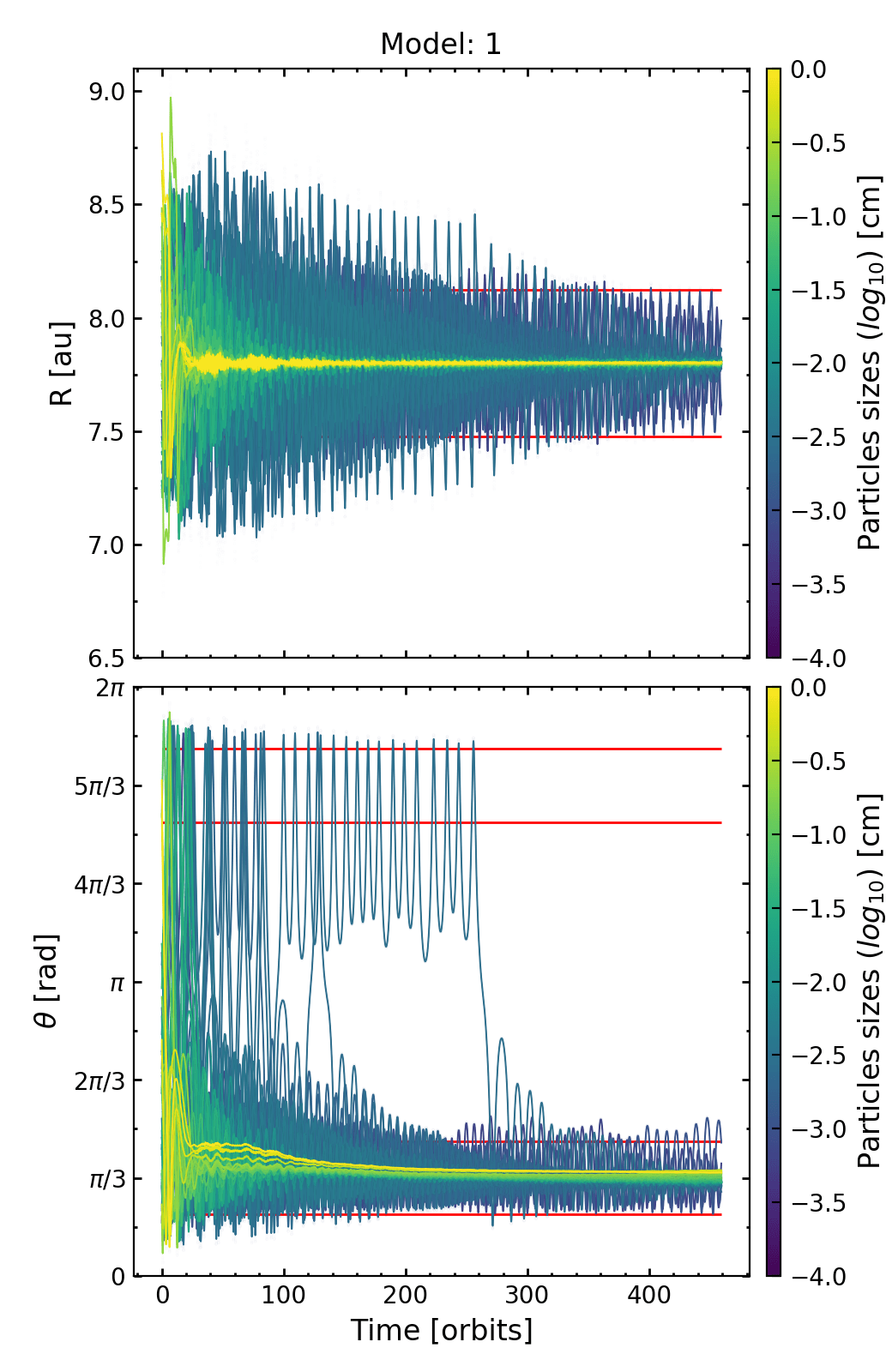}
\caption{Trajectories of some particles ($\mu$m to cm) from model 1 that end up trapped inside the Lagrangian point $L_4$ only. The top horizontal red lines represent the location $L_5$, while the bottom horizontal lines represent $L_4$. Some submm particles oscillate in horseshoe orbits (between both $L_4$ and $L_5$) before ending in $L_4$. Large mm-cm particles oscillate only in tadpole orbits around $L_4$.}
\label{RvsTime_model1}
\end{figure}

We are interested in the libration of particles around $L_4$ and $L_5$, including tadpole and horseshoe trajectories between both locations. We selected several particles from the last output in the size range  $10^{-4} - 10^{0} \rm cm$ that end up trapped inside the libration area of $L_4$ defined by the azimuthal position: $\pi/3 - \pi/8 < \phi  < \pi/3 + \pi/8$, and radius: $r_{\rm p} -  R_{\rm Hill} < r < r_{\rm p} +  R_{\rm Hill}$ (see Section \ref{Lagrangian_points_L4_L5}, Eq. \ref{eqLagrangeRegion1}). Figure \ref{RvsTime_model1} shows the evolution in time of the radial and azimuthal coordinates of each particle. The horizontal red lines in Figure \ref{RvsTime_model1} indicate the defined libration area.

As soon as these particles start their motion, they engage into a damped oscillation mode between the capture region $r_{\rm p} \pm R_{\rm hill} \sim (R_{\rm hill} \sim 0.5 \rm ~ au$  for model 1). The larger the particle size, the larger the damping. Millimetric and centimetric particles become immediately constrained to oscillate very close around $r_{\rm p}$ (under-damped regime). Submm and $\mu$m particles also oscillate in a under-dumping mode but with a larger amplitude around $r_{\rm p}$ (see top panel of Figure \ref{RvsTime_model1}). Regarding the azimuthal coordinate, mm and cm particles oscillate with small amplitude around a full orbit around the star; they quickly reach the equilibrium position at $L_4$ (recall that we selected only particles ending up in $L_4$; see particles inside the horizontal red lines at the bottom panel of Figure \ref{RvsTime_model1}). However, small particles ($\mu$m and submm) librate with larger amplitudes. Some of them become trapped in the other Lagrangian point $L_5$ for a while, 
going back and forth from $L_4$ to $L_5$, before finally ending up in $L_4$. 

In summary, mm-cm particles that end up in $L_4$ oscillate around $L_4$ (tadpole orbits) during the whole simulation with an amplitude defined by the libration amplitude (Eq. \ref{eqLagrangeRegion1}). Submm and $\mu$m particles oscillate around both Lagrangian points (horseshoe orbits). $L_4$ catches more smaller particles ($\mu \rm m$-submm), and $L_5$ efficiently traps larger particles (mm-cm). 

A major result is that all particles that end up at $L_4$ (or at $L_5$), regardless of their size, were initially located inside the libration region defined by the effective capture radius (Eq. \ref{eqLagrangeRegion3}), that is, an annulus centered at $r_{\rm p}$ with a width of $\Delta r = 2 \times R_{\rm Hill}$ ($\sim \rm 1 au$ for model 1). No particles initially located outside this area end up trapped around a Lagrangian point. Particles traveling from outer regions dragged by the gas pass through the gap but without being trapped in tadpole orbits around a Lagrangian point.
The capture region suggests that the mass reservoir to be accumulated in a Lagrangian point should be $M_{\rm res}^{L_4,L_5} = \int_{r_{\rm p}-\Delta r}^{r_{\rm p}+\Delta r} \sigma_{\rm dust}(r) 2 \pi r {\rm d}r$, where $\sigma_{\rm dust}(r)$ is the initial dust density profile, and $r_{\rm p}$ the planet position.

This maximum imposes some constraint on the in situ  formation scenario of Trojan planets. For instance, from the initial condition defined for model 1 (Section \ref{Gas_initialization}, Eq. \ref{eqdens}), we have $\sigma_{\rm dust}(r) = \Sigma_{\rm gas}(r)/100$, which gives $M_{\rm res}^{L_4,L_5} \sim 184$ $M_{\rm moon}$ (2.3 earth masses). For this model, the final masses around $L_4$ and $L_5$ are 2.13 and 3.0 $M_{\rm moon}$, respectively (see Table \ref{TableLpoints}), representing a capture efficiency of about $1.2\%$ ($L_4$) and $1.6\%$ ($L_5$) of $M_{\rm res}^{L_4,L_5}$. Depending on the simulation parameters, the capture efficiency may vary. However, the mass reservoir is an independent constraint, which will not change when varying the parameters or if a larger or smaller disk is assumed.

\subsubsection{Outer disk}

\begin{figure}
\centering
\includegraphics[width=0.9\linewidth]{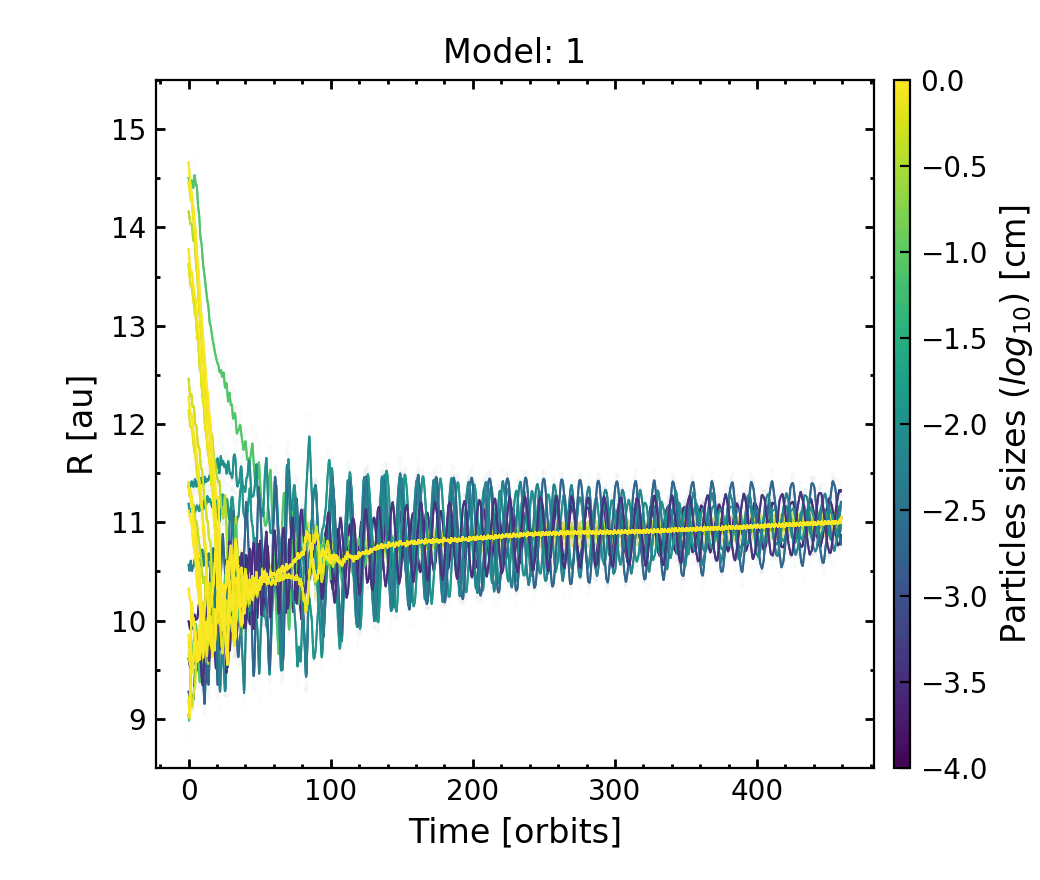}
\caption{Radial trajectory of randomly selected particles ($\mu$m to cm) that end up trapped in the outer ring located at $r = 11$ au featured at the end of the simulation model 1.}
\label{RvsTime_planetX_model1}
\end{figure}

The outer disk shows a wide ring structure located at $r \sim 11$ au composed of $\mu$m and submm particles. Within this ring, a large vortex is produced as a consequence of the gravity of the planet, with an over-density of mm-cm particles with a total mass of 16.6 $M_{\rm moon}$ (bottom-right panel of Figure \ref{vortensity_10mil}).

We selected some of the trapped grains in the vortex to trace their trajectories back in time. Figure \ref{RvsTime_planetX_model1} shows the evolution of those particles starting from $t=0$ to $t=10\,000$ yr. We find that the particles were initially located at radial positions between 8.8 to 15 au (the vortex is located at $r \sim 11$ au). 

Some large (mm-cm) particles migrate outward, driven by forces from planetary wakes, while others migrate inward, influenced by drag. Submm dust particles trapped in the vortex were initially located only in orbits close to its final position inside the vortex, that is, $r \sim 11.5$ au (see Figure \ref{RvsTime_planetX_model1}).

As shown in Figure \ref{RvsTime_planetX_model1}, mm-cm particles barely oscillate in the radial direction while they travel through the disk. In contrast, small, submm particles oscillate with decreasing amplitude in the radial direction while they move toward the vortex.

Analyzing the vortex dynamics as a whole, we find that at $t = 1\,000$ yr, it revolves with a mean motion resonance (MMR) of 3:2 with respect to the planet, shifting radially to its final position at $r \sim 11$ au reaching a MMR of 5:3 at the end of the simulation. The resonant vortex, with its dusty banana-shaped structure (bottom panel of Fig. \ref{vortensity_10mil}), could feature some observational signatures such an asymmetry in the dust continuum of the dusty ring to which it belongs, peaking in scattered infrared light and submm emission (e.g., \citealt{Bae+2016, Baruteau+2019}). However, further radiative analysis is beyond the scope of this work.

In our simulations, the planet was not able to migrate. However, for completeness, we run two dedicated models: (i) one in which a 10 $M_J$ planet `feels' the disk, starting a type II migration regime; (ii) and an identical model, but without migration (see parameter Table \ref{models_parameters}, models 7 and 8). It is worth mentioning that, in general a massive planet migrates faster than a lower mass planet. For instance, at first order, a reference migration timescale would be $\tau_{\rm mig} \propto M_{\rm p}^{-1}$ (e.g., \citealt{ArmitageBook}). In our simulation, a $10\,M_J$ planet starts a fast inward type II migration regime from 5.2 to 3.7 au in 460 orbits.

The effect of planet migration slightly reduces the total mass accumulated in both $L_4$ and $L_5$, but the asymmetry favoring $L_5$ over $L_4$ remains. Migration does not enhance material accumulation in $L_4$ over $L_5$, or trigger any destabilization mechanism around $L_5$ as proposed by \cite{Gomes1998}. The only difference we find is on the trajectories of micron particles coupled to the gas. Those particles are not captured in stable tadpole orbits in the horseshoe region, which is probably because of the large planetary mass  used (needed to obtain a fast type II migration regime).

\subsection{Radio flux from Lagrangian points}

\begin{figure}
\centering
\includegraphics[width=0.9\linewidth]{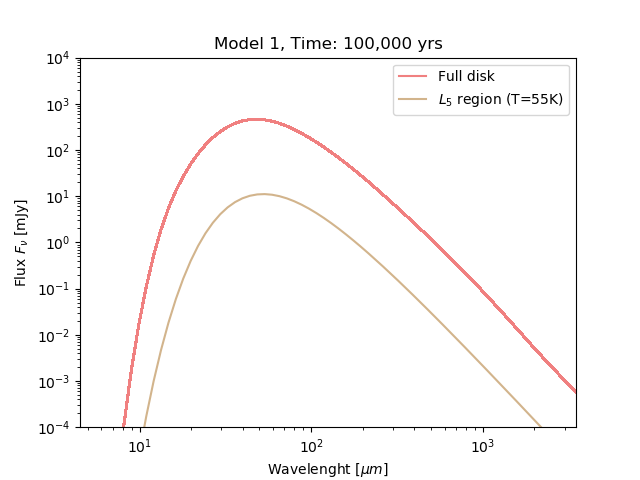}
\caption{Spectral energy distribution from the disk, and the Lagrangian point $L_5$ from model 1 computed at the last evolutionary stage. Emission from the Lagrangian point comes from an optically thin dusty swarm of $\sim 3 M_{\rm moon}$ with a mean temperature of $\sim 55$ K.}
\label{Flux1}
\end{figure}

From the last outputs of the simulation, noting that the system has reached a quasi-stationary regime, we can estimate the emission from the disk and the Lagrangian points. Assuming that the source is located $D = 150$ pc away, the total disk flux can be estimated by integrating the Planck function $B_\nu$ over the disk surface; $F_\nu = (2\pi/D^2) \int_{\rm R_{\rm min}}^{\rm R_{\max}} B_\nu(T(r)) r {\rm d}r$, obtaining a peak flux of about $\sim 450 $ mJy at 50 $\mu$m for model 1. On the other hand, the dust accumulated in a Lagrangian point contributes with a specific localized emission. If it comes from an optically thin region, the emission can be computed from $F_\nu = (1/D^2)  M_{\rm dust} \kappa_\nu B_\nu(T)$, where $M_{\rm dust}$ corresponds to the dust mass located at the Lagrangian point, and $T$ its temperature \citep{Hildebrand1983}. Using the values from Table 1 ($M_{\rm dust} \sim 3 M_{\rm moon}$, $T \sim 55$K), we obtain for model 1 an integrated flux of $\sim 20 $ mJy also peaking at 50 $\mu$m. This is an idealized estimation, without taking noise or instrument limitations   into account. The spectral energy distribution for this model is shown in Figure \ref{Flux1}.

\section{Discussion}\label{discussion}

We studied the evolution of dust present in a viscous disk with an initial gas-to-dust ratio of 100:1. The disk has an embedded approximately Jupiter-mass planet located at 7.8 au which we follow for 460 orbits.
The dust is treated as Lagrangian particles with a full spectrum of sizes ranging from $10^{-4}$ to $10^{0}$ cm. Our simulations are done over two stages: The first one computes the gas dynamics by solving the Navier-Stokes equations, including a nonstationary energy equation for an irradiated disk. The second stage computes the dust dynamics in which dust particles `feel' the gravity from both the star and the planet, and the drag caused by the gaseous disk.

We mainly focus on dust dynamics around Lagrangian points $L_4$ and $L_5$, examining the impact of three parameters that play an important role in the evolution of the dust: the mass of the planet $M_{\rm p}$, the turbulent viscosity of the gas $\alpha$, and the stellar irradiation (to heat the gas) from the star $L_\star$. Some general conclusions, independent of the parameter choice, arise from these models:
Once the planet has carved a gap in the disk, two vortices appear located at the Lagrangian points $L_4$ and $L_5$, revealed by the vortensity minima (Figure \ref{vortensity_10mil}). These minima act as dust traps \citep{Crnkovic-Rubsamen+2015, Surville+2016}, accumulating material in the librating region around $L_4$ and $L_5$.
In all the models, we find that $L_4$ invariably captures a larger number of small ($\mu$m-submm) dust grains than to $L_5$. On the other hand, mm and cm particles are always more abundant in the trailing $L_5$ Lagrangian point. \sout{Big-sized} Larger particles account for more mass; therefore, we always find that at the end of the simulation, $L_5$ has accumulated more mass than $L_4$. Typical values of the mass accumulated around $L_5$ are on the order of 2-3 $M_{\rm moon}$. 

The observed asymmetry applies to all the models, consistent with past theoretical works suggesting that in the presence of drag, the orbits of Trojans around $L_5$ are more stable than those around $L_4$ (e.g. \citealt{Peale1993, Murray1994}). The origin of this asymmetry is due to an over-density created at the trailing region of the planet (e.g., \citealt{Val-Borro+2006, Lyra+2009}).  Such an over-density (equivalent to a pressure bump) in $L_5$ can be understood as follows: the orbital motion of the planet induces a larger gas depletion in the leading direction because the planet excites pressure waves that can remove angular momentum away from Lindblad resonances \citep{GoldreichTremaine1979, Artymowicz+1993}, thus producing  an asymmetry between the density lobes at the leading and trailing zones. This explains, for instance, why in a pressureless or inviscid (no viscosity) model the Lagrangian points  accumulate an equal amount of material. In our case, we show that when lowering the viscosity, the mass ratio $L_4/L_5$ approaches unity. 

A major result concerns the mass reservoir that feeds the Lagrangian points. According to our models, particles ending up trapped either in $L_4$ or $L_5$ at the end of the simulation were initially located only at the co-orbital region of the planet $r_{\rm p}$ within the range $\sim \pm 2 R_{\rm Hill}$ (Fig. \ref{Histograma_Initial_random}). No particles in further orbits, located in the external regions of the disk (i.e., $r > r_p + 2 R_{\rm Hill}$), were trapped in stable orbits by the Lagrangian points (Figure \ref{RvsTime_model1}). This suggests that the total mass available to be trapped in a Lagrangian point of a planet located at $r_{\rm p}$, is present at the beginning of the disk evolution inside the co-orbital area defined by an effective capture range given by Eq.~\eqref{eqLagrangeRegion3}. Therefore, the mass reservoir to feed a Lagrangian point will be: $M_{\rm res}^{L_4,L_5} = \int_{ r_{\rm p} - R_{\rm Hill}}^{ r_{\rm p} + R_{\rm Hill}} \sigma_{\rm dust}(r) 2 \pi r {\rm d}r $, where $\sigma_{\rm dust}(r)$ is the initial dust density distribution. This constraint holds regardless of the free parameters or the radial extension of the disk; and its evaluation is only a function of the initial dust density distribution and the mass of the planet (to compute $R_{\rm Hill}$). 

However, not all this mass is effectively captured by a Lagrangian point. Roughly speaking,  $\sim 1 - 2 \%$ (1-3 $M_{\rm moon}$) of the initial co-rotating dust $M_{\rm res}$ ($\sim 2.3$ earth mass) will end up trapped as Trojan dust. Interestingly, this estimated mass can be taken to be, for instance, the origin of a swarm of material that will produce a Trojan moon-like planet around a Jupiter planet, as suggested by \citet{Beauge+2007}.

The lack of initial material in the co-orbital zone and the dust capturing percentage ($\sim 1-2\%$) in $L_4$ or $L_5$ makes it challenging to assemble a co-orbiting earth-size planet in a Lagrangian point from a primordial configuration of the protoplanetary disk (e.g., assuming a minimum mass solar nebula model \citealt{Weidenschilling+1977, Desch2007}) unless a very thick and massive initial dusty disk with larger grains is considered. For instance, based on the parameters of model 1 (one Jupiter mass planet located at 7.8 au) and assuming a dust capturing efficiency of 1\%, in order to assemble one Earth mass in a Lagrangian point we need an initial dust density value of about $\sigma_L^{\rm dust} = 54 \, {\rm g\,cm^{-2}}$ at that Lagrangian point. This value is consistent with the disk density model presented by  \citealt{Lyra+2009}.

Recent observations of Jovian Trojans in the Solar System indicate a large number of objects\footnote{\url{https://www.minorplanetcenter.net/iau/lists/JupiterTrojans.html}} located in $L_4$ than in $L_5$ \citep{Yoshida+2005, Nakamura+2008, Pitjeva+2020}. In contrast, our results show that the trailing ($L_5$) point is more efficient at accumulating large grains.

We suggest that in the early evolution of a planetary system, in the pre-transitional phase, Trojans form in situ on short timescales ($\sim 10,000$ yrs) as a consequence of a massive planet. Hence, the chemical composition of Trojans located at $L_4$ or $L_5$ should be similar. Also,  far away asteroids, such as those from the main belt in our Solar System, are probably built from different shapes and components than Trojans. Spectroscopic and photometric observations seem to suggest this, as most Jovian Trojans are D- or P-type, while those from the main belt are C- and S-type \citep{Hartmann+1987, Fitzsimmons+1994}. However, as the system evolved, they were probably contaminated by collisions, heavy bombardment, and modified by 
gravitational interactions with other planets (e.g., \citealt{Pal+2004, Freistetter2006}), as suggested by the Nice model \citep{Tsiganis+2005}.

We also propose that if a protoplanet is found in a transitional disk, for example inside the gap of the circumstellar disk around HD100546, theorized to be the consequence of a Jupiter-mass planet, similar to our model parameters \citep{Bouwman+2003, Tatulli+2011}, two asymmetrical swarms of dust should be located at $L_4$ and $L_5$.  The detection of such swarms could be used to reinforce or infer the presence of an embedded putative planet.

The planetary system around PDS 70 constitutes another interesting astronomical laboratory because recent near-infrared SPHERE and NACO observations revealed the presence of a couple of planets within the disk cavity: PDS~70b \citep{Keppler+2018} and PDS~70c \citep{Isella+2019}. The latter was discovered through $H\alpha$ emission with MagAO and MUSE. Our model suggests that such planetary companions, particularly PDS 70b, should have gathered potentially observable dusty swarms in their Lagrangian points. However, a more detailed study of the dynamical interaction between PDS 70b and PDS 70c is required in order to give an accurate prediction.

In our in situ scenario, intrinsic parameters ($M_{\rm p}$, $\alpha$, $L_\star$) play an important role in the primordial structure of  Trojans, especially concerning their reported mass asymmetry. We briefly summarize these effects below. 

\subsection{Viscosity}

Increasing the turbulent $\alpha$ viscosity of the disk enhances the accretion rate (by promoting angular momentum transport), making the dust trapping around the Lagrangian points more difficult. This result can be understood as follows: a high accretion rate means higher values of the radial velocity, producing a reduced potential vortensity at the Lagrangian points to act as attractors. 
Comparing models 1, 2, and 6 (Table \ref{models_parameters}) reflects the mentioned influence of the viscosity (Figure \ref{L4_L5_mass_evolution_model1} next to Figure \ref{L4_L5_mass_evolution_model2_new}). The lower the $\alpha$, the more similar the mass in $L_4$ is to that in $L_5$. 

\subsection{Planetary mass}

Increasing the mass of the planet does not help to increase dust accumulation around a Lagrangian point. For instance, Models 3 and 4 (Figure \ref{L4_L5_mass_evolution_model3_new}) possess a larger planetary mass (5 $M_J$), which more easily destroys the leading swarm in $L_4$ by removing its angular momentum through pressure waves from the massive planet \citep{GoldreichTremaine1979, Artymowicz+1993}. 

\subsection{Stellar irradiation}

Stellar irradiation that heats gas also plays an important role in the final structure around the Lagrangian point. The instability of a vortex begins when the vorticity at its center is close to the background vorticity (e.g., \cite{Surville+2016}). A hotter disk would have a higher background pressure, reducing the difference (gradient) from the background to the center of the  vortex, thus reducing the strength of the vortex. This is observed in model 3 ($5 L_\odot$), where an early instability appears when compared to model 4 ($1 L_\odot$).

\subsection{Final remarks}

We do not include the back-reaction from dust on gas, which can be relevant when the local gas-to-dust ratio is about $\sim 1:1$. In our simulations, we show that at the Lagrangian points, the gas-to-dust ratio could reach values of $\sim 2:1$ (Table \ref{TableLpoints}). Also, in model 1, a large vortex at the outer disk accumulates a considerable amount of dust, reaching a gas-to-dust ratio of $\sim$ 9:1  (Figure \ref{all_density_model1}).

The inclusion of the back-reaction of dust should include a term of the form $\epsilon \frac{\Omega(r)}{St} \Delta \textbf{v}$ (similar to the drag force in Eq.~\ref{eqFdrag}), where $\epsilon$ is the dust-to-gas ratio.  Two-dimensional numerical simulations show that this feedback could reduce the lifetime of vortices when dust density becomes comparable to gas density ($\epsilon \sim 1$) within the vortex, reducing dust trapping \citep{Fu+2014}. However, when the dust density diminishes, the vortex can appear to decrease again due to the growing dust-to-gas ratio. This ensures efficient dust trapping over the  lifetime of the disk \citep{Raettig+2015}. Therefore, with or without a dust back-reaction, this should not affect our conservative estimation of 1-2\% of the mass reservoir ($M_{\rm res}^{L_4,L_5}$) being piled up around Lagrangian points. Also, for a long-term evolution model (1000 orbits), the vortices at the Lagrangian points disappear. However, the dust accumulated during the first stages remains trapped, making the in situ gravitational collapse of dust to form  km-sized asteroids (or planetesimals) at those locations a plausible scenario.

\section{Conclusions}\label{conclusions}

We highlight our main findings as follows:

\begin{itemize}

    \item When a planet carves a gap, it creates an overdensity in the trailing position ($L_5$) compared to the leading one ($L_4$), producing a large vortex at the former (larger pressure bump in $L_5$). The asymmetry is due to pressure waves from the planet and the action of viscosity, which together promote a slightly larger loss of angular momentum at $L_4$. This translates to lower density and pressure around $L_4$ with respect to $L_5$.

    \item $L_4$ accumulates more $\mu$m and submm particles, while $L_5$ efficiently traps larger grains (mm-cm). The total bulk mass is retained in the largest size particles, making $L_5$ always more massive.

    \item Most of the particles trapped in $L_4$ possess Stokes numbers  ${\rm St} \textless 0.1$, while particles trapped around $L_5$ have ${\rm St} \textgreater 0.1$ up to 10.

    \item Planet migration does not appear to influence the observed asymmetry between $L_4$ and $L_5$. However, the free parameters of the model may affect it: lower viscosity tends to $L_4 \sim L_5$. Lower stellar irradiation (colder disks) tends to enhance dust accumulation in both $L_4$ and $L_5$. Larger planetary mass tends to destabilize both, but especially $L_4$ by enlarging the planetary gap. 

    \item The mass reservoir around a Lagrangian point is limited to $M_{\rm res}^{L_4,L_5} = \int_{ r_{\rm p} - R_{\rm Hill}}^{ r_{\rm p} + R_{\rm Hill}} \sigma_{\rm dust}(r) 2 \pi r {\rm d}r $, where $\sigma_{\rm dust}(r)$ is the initial dust density distribution. The retained dust was initially located only at the co-rotating orbital path of the planet, inside the effective capture radius defined by Eq. \ref{eqLagrangeRegion3}.
    
    \item If Trojans appear to have formed at an early evolutionary stage of the solar nebula on short timescales ($10^5$ yr), after which their chemical composition should have been similar. However, their subsequent evolution and composition may have been significantly altered by dynamical instabilities and interactions (as in the Nice model for instance).
    
\end{itemize}

Future observations of embedded planets in disks ---in particular the hypothetical detection of thermal emission in co-rotation with the planet--- will allow us to quantify the role of gas effects for dust trapping around Lagrangian points. This mechanism has profound implications for the formation of Trojans in our Solar System but also in extra-solar systems.

\begin{acknowledgements}
%% -------------------------
MM acknowledges financial support from the Chinese Academy of Sciences (CAS) through a CAS-CONICYT Postdoctoral Fellowship administered by the CAS South America Center for Astronomy (CASSACA) in Santiago, Chile. JO acknowledges financial support from the Universidad de Valpara\'iso and from Fondecyt (grant 1180395). CAG acknowledges Mulatona Cluster from CCAD-UNC, which is part of SNCAD-MinCyT, Argentina. AB acknowledges support from FONDECYT Regular 1190748. NC acknowledges support from the European Union's Horizon 2020 research and innovation programme under the Marie Sk\l{}odowska-Curie grant agreement No 210021. MM, JG, JC, JO, AB, and MS acknowledge support from Iniciativa Cient\'ifica Milenio via the N\'ucleo Milenio de Formaci\'on Planetaria.  
\end{acknowledgements}

%%%%%%%%%%%%%%%%%%%% REFERENCES %%%%%%%%%%%%%%%%%%
    
\bibliographystyle{aa}
\bibliography{astro}

\begin{appendix}

\section{Figures models: 2, 3, 4, and 6}

\begin{figure*}[]
\centering
\includegraphics[width=0.8\linewidth]{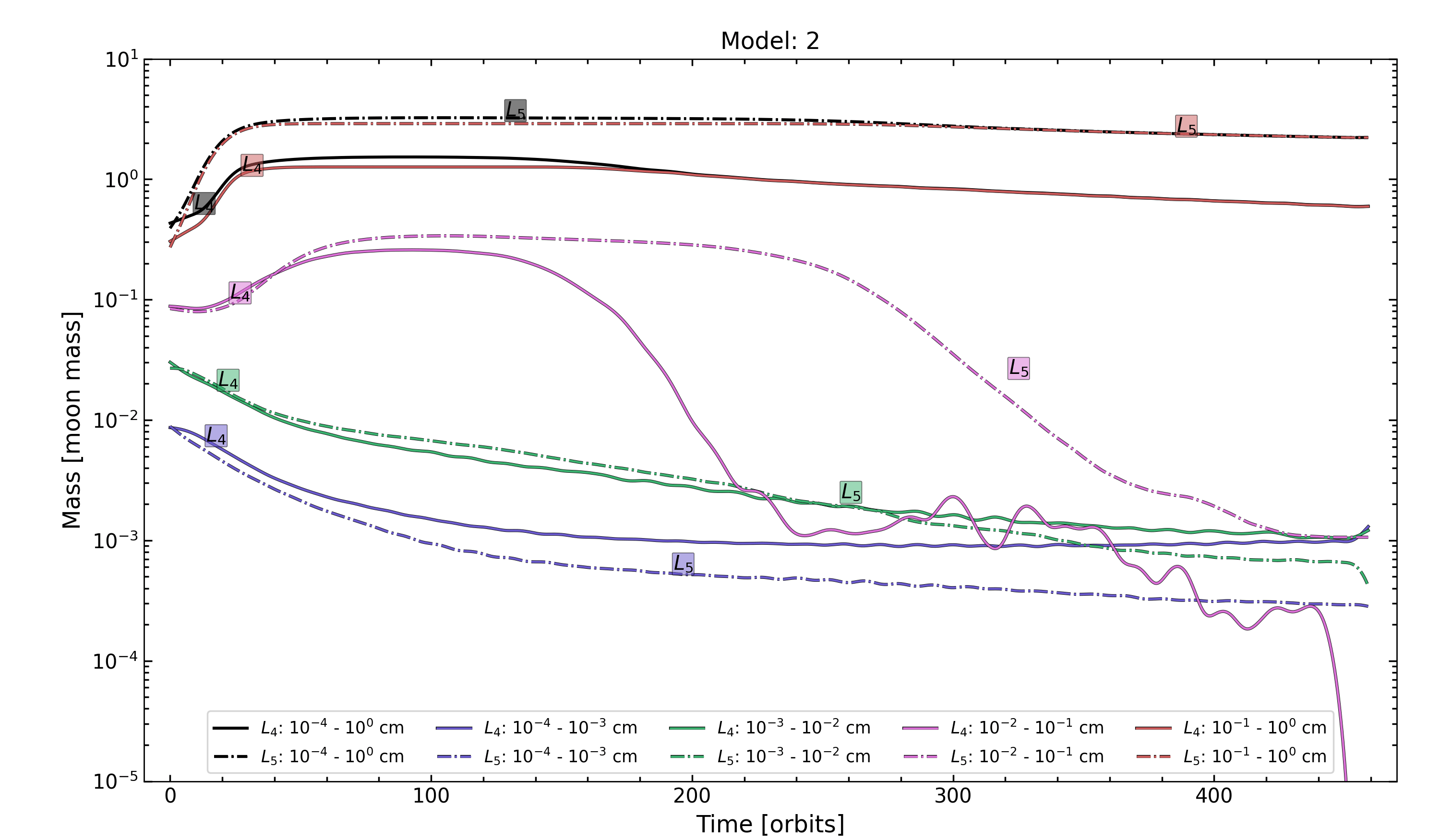}
\caption{Dust mass accumulated over time around the Lagrangian points $L_4$ (continuous line) and $L_5$ (dotted line) for model 2. The evolution is shown
for different sizes, including the effective (all sizes) accumulated mass (black continuous line).}
\label{L4_L5_mass_evolution_model2_new}
\end{figure*}

\begin{figure*}
\centering
\includegraphics[width=0.8\linewidth]{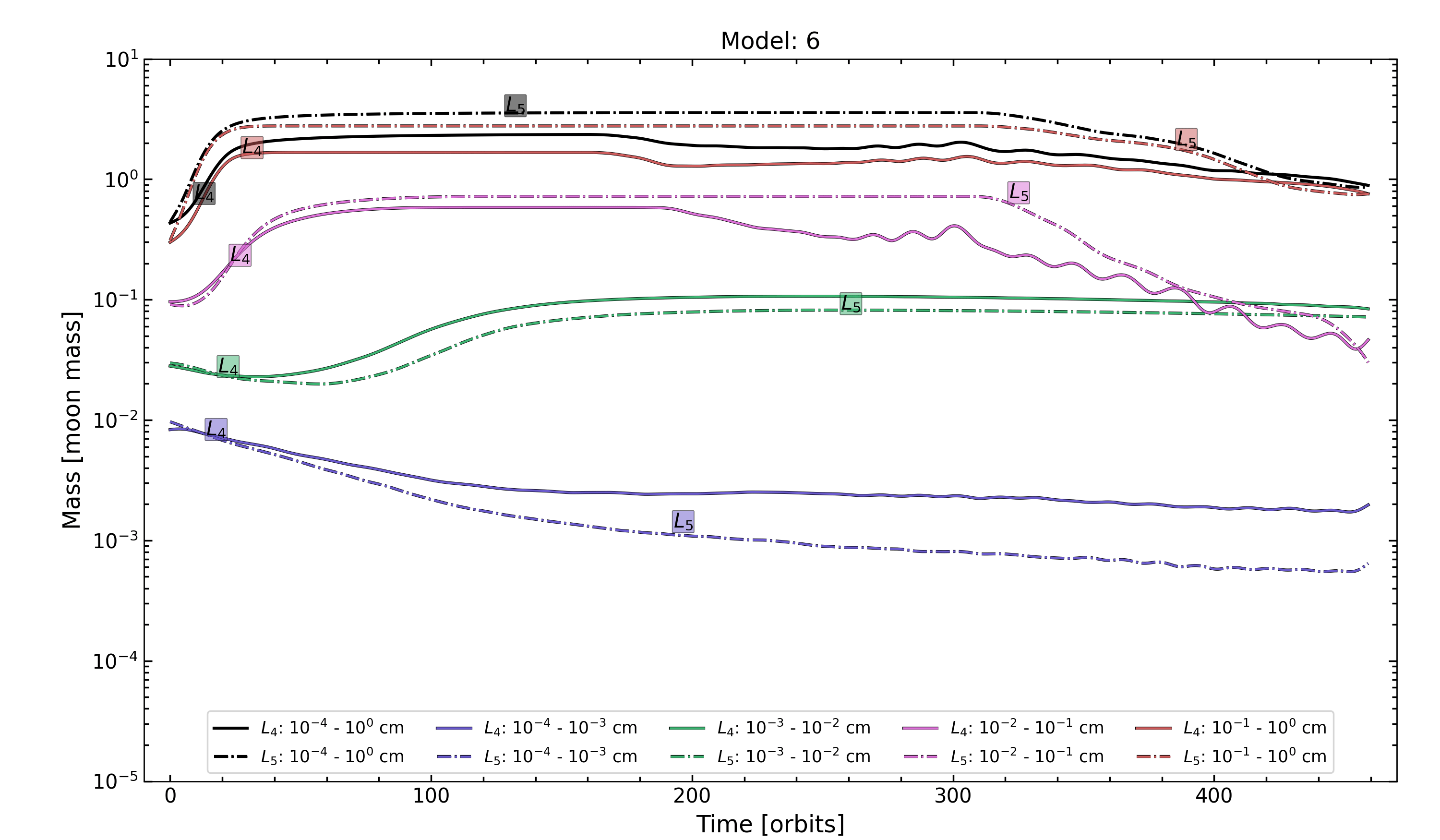}
\caption{Dust mass accumulated in time around the Lagrangian points $L_4$ (continuous line) and $L_5$ (dotted line) for model 6. Evolution is shown
for different sizes, including the effective (all sizes) accumulated mass (black continuous line).}
\label{L4_L5_mass_evolution_model6_new}
\end{figure*}

\begin{figure*}[h!]
\centering
\includegraphics[width=0.8\linewidth]{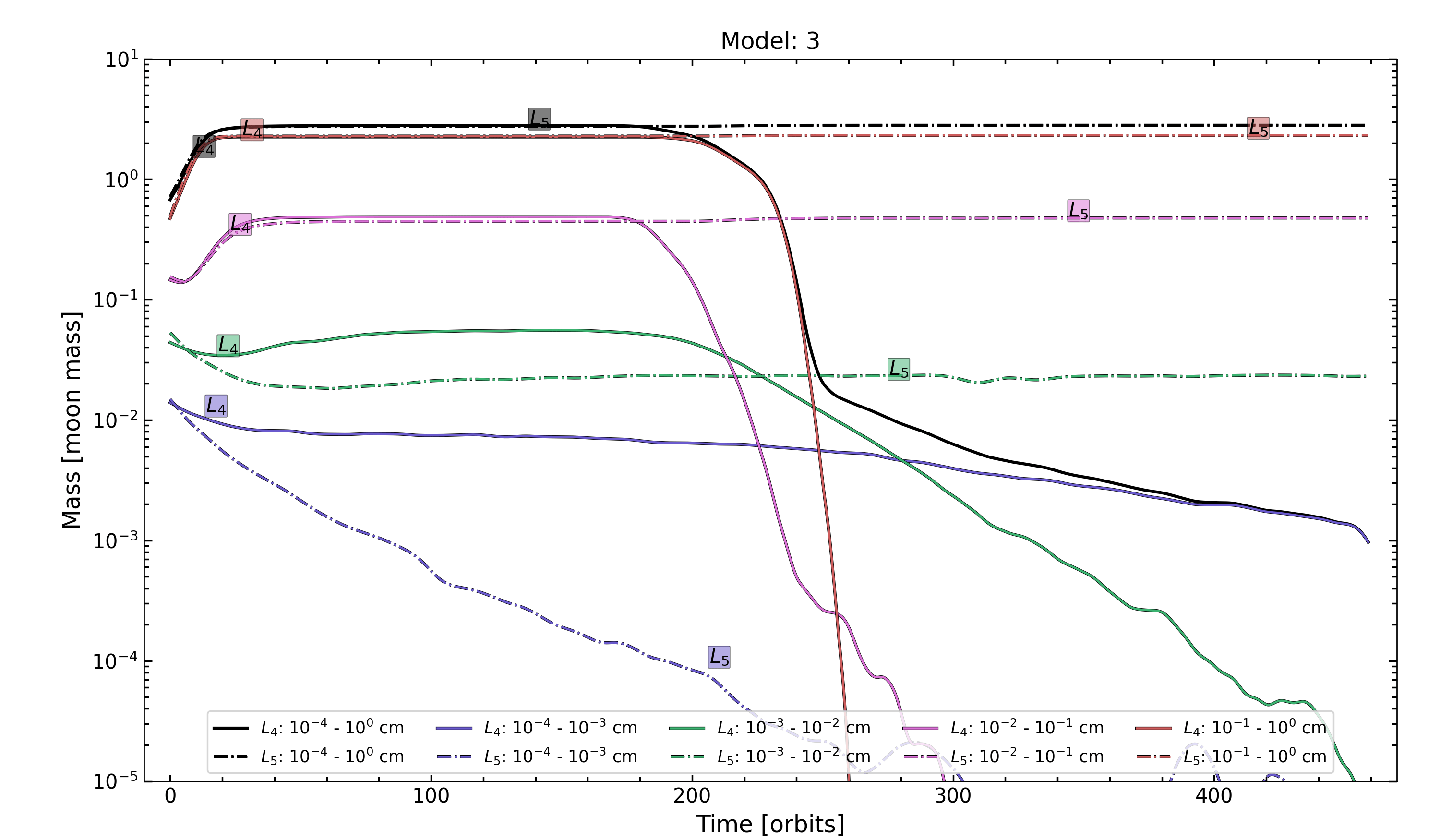}
\caption{Dust mass accumulated over time around the Lagrangian points $L_4$ (continuous line) and $L_5$ (dotted line) for model 3. Evolution is shown for different sizes, including the effective (all sizes) accumulated mass.}
\label{L4_L5_mass_evolution_model3_new}
\end{figure*}

\begin{figure*}
\centering
\includegraphics[width=0.8\linewidth]{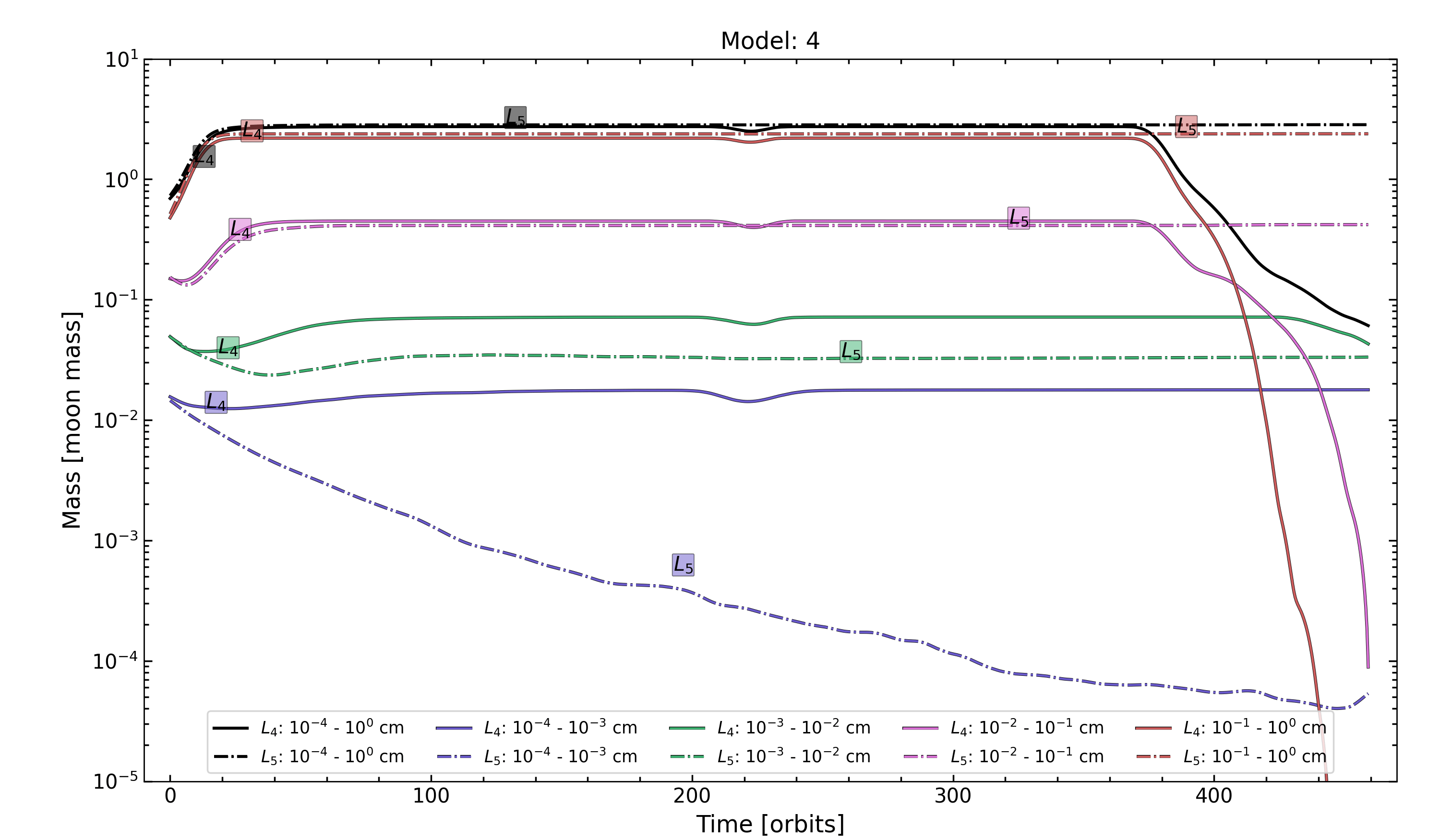}
\caption{Dust mass accumulated over time around the Lagrangian points $L_4$ (continuous line) and $L_5$ (dotted line) for model 4. Evolution is shown for different sizes, including the effective (all sizes) accumulated mass.}
\label{L4_L5_mass_evolution_model4_new}
\end{figure*}

\begin{figure}
\centering
\includegraphics[width=0.8\linewidth]{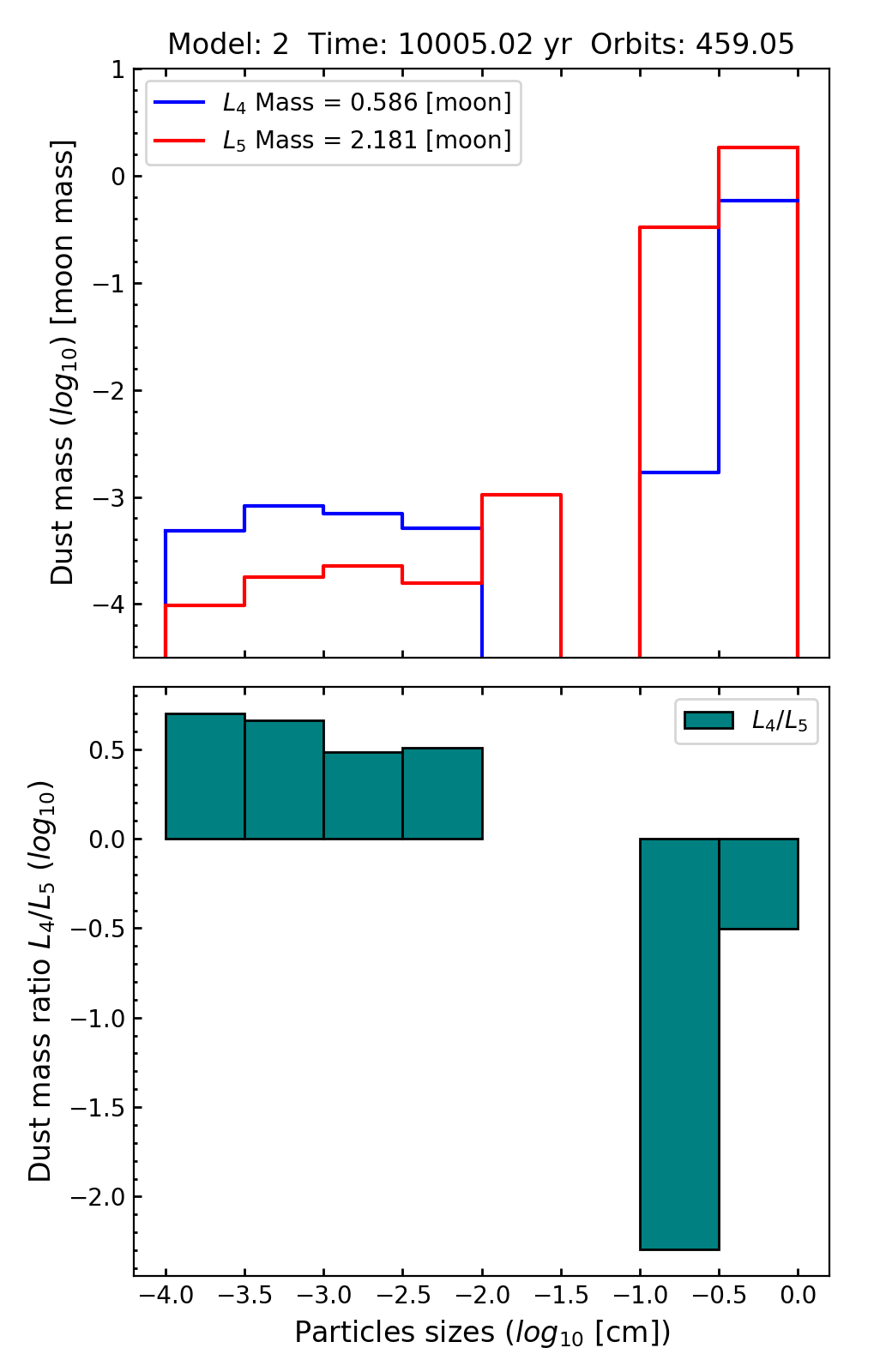}
\caption{Mass spectrum of model 2 as a function of particle size for the dust accumulated in $L_4$ and $L_5$ for the last evolutionary stage (460 orbits). The bottom panel shows the mass ratio $L_4/L_5$ as a function of particle size.}
\label{histo_L4_L5_model2}
\end{figure}

For a better visualization we regroup all the figures concerning models 2 (Figs \ref{L4_L5_mass_evolution_model2_new}, \ref{histo_L4_L5_model2}), 3 (Fig. \ref{L4_L5_mass_evolution_model3_new}), 4 (Fig. \ref{L4_L5_mass_evolution_model4_new}), and 6 (Fig. \ref{L4_L5_mass_evolution_model6_new}) in this Appendix.

\begin{figure}
\centering
\includegraphics[width=0.8\linewidth]{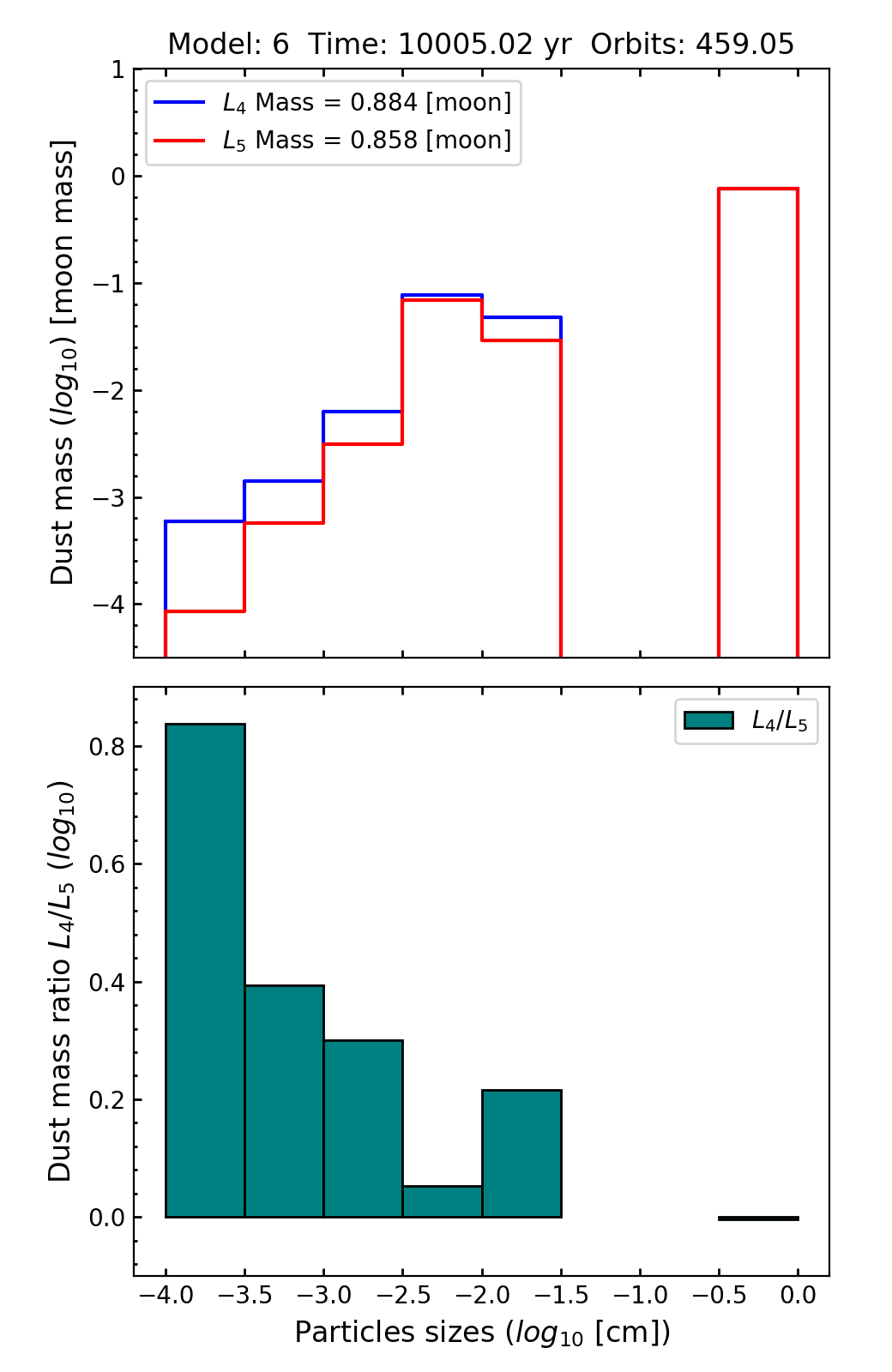}
\caption{Mass spectrum of model 6 as a function of particle size for the dust accumulated in $L_4$ and $L_5$ for the last evolutionary stage (460 orbits). The bottom panel shows the mass ratio $L_4/L_5$ as a function of particle size.}
\label{histo_L4_L5_model6}
\end{figure}

\end{appendix}

% Don't change these lines
%\bsp   % typesetting comment
%\label{lastpage}
\end{document}